\documentclass[trackchanges,author-year]{aastex701}
\usepackage{amsmath}
\usepackage{booktabs}
\usepackage{multirow}
\usepackage{graphicx}
\usepackage{sidecap}
\usepackage{subcaption}
\usepackage[section]{placeins}

\begin{document}

\title{The Role of Intrinsic Temperature and Vertical Mixing in Characterizing Sub-Neptune Atmospheres.}

\author[orcid=0000-0002-5977-9317,sname='Dushyantha Kumar']{Neha Dushyantha Kumar}
\affiliation{Department of Astronomy \& Astrophysics, 525 Davey Laboratory, The Pennsylvania State University, University Park, PA 16802, USA}
\affiliation{Center for Exoplanets and Habitable Worlds, The Pennsylvania State University, University Park, PA 16802, USA}
\email[show]{nehadushyanth@gmail.com}  

\author[0000-0002-2990-7613]{Jessica E. Libby-Roberts}
\affil{Department of Astronomy \& Astrophysics, 525 Davey Laboratory, The Pennsylvania State University, University Park, PA 16802, USA}
\affil{Center for Exoplanets and Habitable Worlds, The Pennsylvania State University, University Park, PA 16802, USA}
\affil{Department of Physics and Astronomy, University of Tampa, Tampa, FL 34606, USA}
\email{}

\author[0000-0003-4835-0619]{Caleb I. Ca\~nas}
\affiliation{Southeastern Universities Research Association, Washington, DC 20005, USA}
\affiliation{NASA Goddard Space Flight Center, 8800 Greenbelt Road, Greenbelt, MD 20771, USA}
\email{c.canas@nasa.gov}

\author[orcid=0000-0002-0413-3308] {Nicholas F. Wogan}
\affiliation{Space Science Division, NASA Ames Research Center, Moffett Field, CA 94035, USA}
\affiliation{Virtual Planetary Laboratory, University of Washington, Seattle, WA 98195, USA}
\email{...}

\author[0000-0001-9596-7983]{Suvrath Mahadevan}
\affil{Department of Astronomy \& Astrophysics, 525 Davey Laboratory, The Pennsylvania State University, University Park, PA 16802, USA}
\affil{Center for Exoplanets and Habitable Worlds, The Pennsylvania State University, University Park, PA 16802, USA}
\email{}

\author[0000-0003-1622-1302] {Sagnick Mukherjee}
\affiliation{School of Earth and Space Exploration, Arizona State University, Tempe, AZ 85287, USA}
\email{smukhe50@asu.edu}

\begin{abstract}

Sub-Neptune planets are often modeled with a dense rocky or metal-rich interior beneath a thick hydrogen/helium (H/He) atmosphere; though their bulk densities could also be explained by a water-rich interior with a thin H/He atmosphere. Atmospheric composition provides a key mechanism to break this degeneracy between competing interior models. However, the overall composition of sub-Neptunes inferred from spectra obtained with the James Webb Space Telescope, remains debated in part due to differences in modeling assumptions. While previous studies explored parameter spaces such as stellar spectra, atmospheric metallicities, and carbon-to-oxygen ratios, they often assumed fixed intrinsic temperatures ($T_{\mathrm{int}}$) and vertical eddy diffusion coefficients ($K_{\mathrm{zz}}$) - two critical, yet poorly constrained, drivers of atmospheric chemistry. To address this, we present a self-consistent grid of models that covers the full plausible range of $T_{\mathrm{int}}$ (60 - 450 K) and $K_{\mathrm{zz}}$ ($10^{5}$ - $10^{12}$ cm$^2$/s) using the open-source \texttt{PICASO} and \texttt{VULCAN} packages to better characterize sub-Neptune atmospheres. Focusing on K2-18b analogs, we demonstrate that $T_{\mathrm{int}}$ and $K_{\mathrm{zz}}$ significantly impact CH$_4$, CO$_2$, CO, NH$_3$ and HCN abundances, with H$_2$O being largely unaffected. Our work demonstrates that comprehensive parameter space exploration of thermal and mixing parameters is essential for accurate interpretation of sub-Neptune spectra, and that single-parameter assumptions can lead to misclassification of planetary interiors. We provide a diagnostic framework using multi-molecule observations to distinguish between competing atmospheric models and advance robust characterization of sub-Neptunes.

\end{abstract}


\section{Introduction}\label{Introduction} 

Sub-Neptunes, exoplanets with radii between $1.6$ and $3.2\,R_{\oplus}$ \citep[e.g.][]{Rogers2015,Wolfgang2016,Fulton2017} are one of the most common planet populations discovered so far \citep[e.g.][]{Fulton2018,Hsu2019}. Without sub-Neptune analogs in our own Solar System, we must rely on bulk densities to inform our understanding of their interior compositions. One explanation of their bulk densities is a smaller, denser interior beneath a thick hydrogen/helium ($\text{H}/\text{He}$) envelope \citep{Rogers2010a, Rogers2010b,Benneke2013}. However, their bulk densities of 1--3 g cm$^{-3}$ (compared to Earth's $5.5$ g cm$^{-3}$) can also be explained by interiors comprised of deep, global water oceans beneath a thin H/He atmosphere, often denoted as Hycean worlds \citep{Madhusudhan2021}. Alternatively, these densities can be accounted for by soot planets, which incorporate refractory organic carbon (referred to as soot) as a major component \citep{li2025}. Sub-Neptunes, specifically Hycean worlds \citep{Madhusudhan2021}, are a subject of detailed habitability studies \citep[e.g.][]{Madhusudhan2023, Innes2023, Mitchell2025}. We can break the degeneracy in a planet's bulk density by characterizing their fundamental differences which lie in their atmospheric composition \citep[e.g.,][]{Madhusudhan2021, Yu2021, Tsai2021, Benneke2024, Wogan2024, Cooke2024, li2025}. 

Exoplanet atmospheric characterization primarily uses transmission and emission spectroscopy, with missions like the James Webb Space Telescope (JWST) \citep{Gardner2006} enabling new levels of precision \citep{Espinoza2025, deWit2025}. However, interpretation of these observations require atmospheric models, which broadly fall into two categories: self-consistent forward modeling and retrievals \citep{Madhusudhan2009, Howe2017}.

Self-consistent forward modeling involves constructing atmospheric models based on a realistic combination of underlying physical and chemical processes (e.g., radiative-convective equilibrium, thermochemical equilibrium, photochemistry) that self-consistently predict temperature-pressure profiles, chemical abundances, and synthetic spectra from first principles, which can then be directly compared to observations using various statistical metrics \citep{Madhusudhan2009, Howe2017}. Retrievals, by contrast, statistically invert observed spectra to directly infer atmospheric parameters, offering flexibility to explore regions of parameter space not constrained by forward model assumptions (with the danger that some solutions may be nonphysical). This approach, therefore, can yield more comprehensive uncertainty estimates and identify potential model inadequacies \citep{Kreidberg2018}. 

Atmospheric dynamics is complex \citep{Deming2017, Roman2019, Steinrueck2023}, dependent on both planetary and stellar factors. Key model influences include large scale inputs such as stellar irradiation \citep{Fortney2018} and planetary masses, radii, and equilibrium temperatures, down to the smaller scale processes of chemical reactions \citep{Visscher2010, Moses2011}, 3D atmospheric dynamics, and molecular line lists \citep{Marley2021}. Because these factors are complex and at times not well-constrained, interpretations of exoplanet atmospheric models are highly sensitive to the choice of underlying assumptions \citep{Barstow2017}.

Metallicity, carbon-to-oxygen (C/O) ratios and stellar luminosity (in the Ultraviolet) driven photochemistry have been the dominant parameters explored in sub-Neptune atmospheric models \citep{Shorttle2024, Hu2025, Jaziri2025, Mukherjee2025, Crossfield2025}. \citet{Yu2021} extended the parameter space to explore the impacts of shallow versus deep surfaces. However, the majority of these studies assume a constant intrinsic temperature ($T_{\mathrm{int}}$) and vertical mixing coefficient ($K_{\mathrm{zz}}$) throughout the model grid; values that, although based on educated guesses, remain poorly constrained by observations. Carbon, oxygen and nitrogen-bearing species are central to atmospheric models because they drive disequilibrium chemistry via quenching, thus determining the deep atmosphere quench point that controls the observable abundances of key molecules like $\text{CH}_4$ and $\text{CO}$ \citep{Zamyatina2024}. Recent JWST observations of the warm-Neptune, WASP-107b  \citep{Sing2024, Welbanks2024}, demonstrate that vigorous $K_{\mathrm{zz}}$ and a hot intrinsic temperature $T_{\mathrm{int}}$ are required to explain order‐of‐magnitude $\text{CH}_4$ depletion and CO enhancement. Similarly, disequilibrium retrievals highlight the sensitivity of species like NH$_3$ to vertical transport and interior heat \citep{Soni2024, Hu2025}. 

Moreover, photochemical recycling in sub-Neptune atmospheres depends critically on both interior temperature and surface depth. In deep, surface-less envelopes, CH$_4$ and NH$_3$ produced in the upper atmosphere can be carried downward, converted back to equilibrium species in hot layers, and then mixed upward again, preserving their abundances against photolytic loss. By contrast, a shallow surface ($\leq 10\,$bar) cuts off this recycling: CH$_4$ and NH$_3$ are destroyed by stellar UV and cannot be replenished, shifting the chemistry toward CO, CO$_2$, and N$_2$ dominance \citep{Yu2021}. Thus, both hot interiors and shallow surfaces produce low CH$_4$ but through different mechanisms. Water condensation at the lower boundary further depletes oxygen-bearing molecules, raising the C/O ratio and promoting higher-order hydrocarbons and organic hazes \citep{Huang2024}, though existing chemical networks may underestimate destruction of complex organics \citep{Tsai2021}. 

A significant limitation in previous investigations is that models used to interpret disequilibrium chemistry rely on simplified pressure-temperature ($P$-$T$) profiles. This leads to atmospheric chemistry not being fully self-consistent with radiative-convective transport. To expand on earlier work, we systematically explore the previously under constrained parameter space of intrinsic temperature and vertical mixing to characterize their influence on sub-Neptune atmospheric composition and observability. We create a set of model grids that couple hot interior structure with quench driven chemistry for a K2-18b sub-Neptune analog using the radiative-convective code \texttt{PICASO} \citep{Batalha2019, Batalha2022, Mukherjee2023} and the photochemical code \texttt{VULCAN} \citep{Tsai2021} (Section \ref{sec:Methods}). Our model grid explores $T_{\mathrm{int}}$ ranging from $60$--$450\,\mathrm{K}$ and $K_{\mathrm{zz}}$ covering $10^{5}$ - $10^{12}$ cm$^2$/s, along with the variable $K_{\mathrm{zz}}$ profile specified in \citet{Hu2021, Wogan2024}, thus spanning the full range of plausible, yet unexplored, mixing and thermal states. For each case, we compute transmission spectra from $0.6$--$5\,\mu\mathrm{m}$, analyzing CH$_4$, CO, CO$_2$, and NH$_3$ features. 

From this comprehensive parameter space exploration, we present our key results in Section \ref{sec:Results} by examining molecular abundance patterns that reveal distinct sensitivity regimes for different species (Subsection \ref{sec:Tint+Kzz Effects}), vertical abundance profiles that demonstrate quenching behavior across temperature and mixing conditions (Subsection \ref{sec:Quenching}), and transmission spectra that show systematic spectral evolution reproducing observed K2-18b features (Subsection \ref{sec: spectra}). In Section \ref{sec: Discussion}, we discuss the implications our results have on the current population of sub-Neptunes, and we conclude in Section \ref{sec:conclusion} by summarizing our diagnostic framework for distinguishing between competing atmospheric scenarios and its broader applications to temperate sub-Neptune characterization with current and future observatories.

\section{Methods} \label{sec:Methods}

\subsection{Target Selection: K2-18b Analog}\label{subsec: Target Characterization}

In this work, we define K2-18b analogs as temperate sub-Neptunes with radii, masses, and equilibrium temperatures similar to  K2-18b (Table \ref{tab:combined-params}), where both Hycean world and gas-rich sub-Neptune interpretations remain plausible. K2-18b \citep{Montet2015}, with its well-constrained radius and mass \citep[$2.610 \pm 0.087,R_{\oplus}$, $8.63 \pm 1.35,M_{\oplus}$, respectively;][]{Benneke2019}, serves as an ideal test case to distinguish how interior heat and mixing influence the observed spectra. Existing atmospheric analyses span a wide range of proposed compositions, from water-rich Hycean worlds with relatively thin H/He atmospheres \citep{Madhusudhan2021, Madhusudhan2023, Hu2025} to gas-rich sub-Neptunes with thick hydrogen envelopes \citep{Wogan2024, Shorttle2024, Madhusudhan2020, Benneke2019}. Independent reanalyses and joint retrievals have demonstrated that key inferences, such as CO$_2$ strength and the presence of sulfur species, can be sensitive to data reduction and modeling assumptions \citep{Cooke2024}. 

These ongoing investigations of K2-18b have revealed the complexity of distinguishing between different atmospheric scenarios for sub-Neptune planets, motivating us to explore additional regions of parameter space that have been previously under-explored. By systematically varying the intrinsic temperature and vertical mixing - parameters typically held fixed in earlier modeling efforts - we aim to understand how these critical drivers influence atmospheric chemistry and observable signatures in K2-18b analogs. We compare our model grid predictions with existing atmospheric models for K2-18b, particularly the sub-Neptune framework developed by \citet{Wogan2024}, to validate our approach and demonstrate how thermal and mixing parameters affect molecular abundance patterns. This comprehensive parameter space exploration will enable future studies to better disentangle the complexities of sub-Neptune atmospheric composition by providing a systematic framework for understanding how interior thermal structure and vertical transport processes control observable spectral features across the broader population of temperate sub-Neptunes.

\begin{figure}[!h]            
  \centering                    
  \includegraphics[width=\linewidth        
  ]{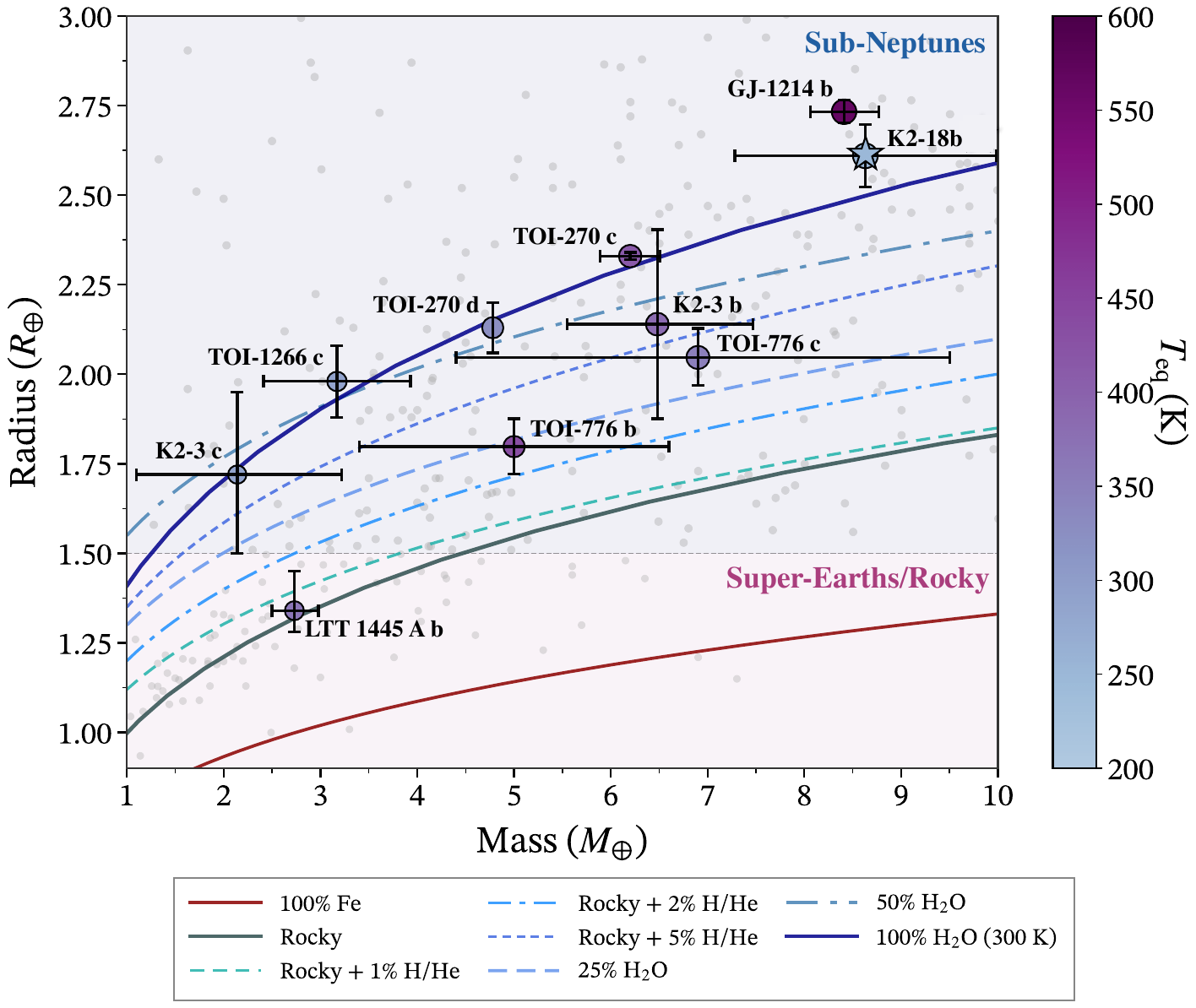}              
  \caption{Transiting exoplanets with measured masses and radii from the \texttt{exoatlas} \textit{Transiting Exoplanets} catalog (\citealt{Berta-Thompson2025}). The gray points represent the sub-neptune population extracted from \texttt{exoatlas}. The selected possible Hycean worlds from \citet{Madhusudhan2021} are labeled with error bars for reference; point color encodes equilibrium temperature (K) ranging from 200 K to 600 K using values taken from the NASA Exoplanet Archive (\citealt{Christiansen2025}) assuming zero albedo. K2-18b analogs -- sub-Neptunes with comparable bulk properties and equilibrium temperatures (250-300 K) that occupy the compositional transition region where both Hycean world and gas-rich sub-Neptune interpretations remain plausible, are highlighted within this population. The regime boundaries between super-Earth/rocky composition (light purple) and sub-Neptune/gas-rich (blue) are based on the radius criterion of $1.6 R_{\oplus}$ from \citet{Rogers2015}, with sub-Neptunes defined as planets with radii between 1.6 and $3.2 R_{\oplus}$. Solid and dashed composition curves from \citet{Zeng2008,Zeng2019} for rocky and hydrogen-rich planets showcase the degeneracy in mass-radius space. Shaded regions are not strict classifications.}
  \label{fig:mass-radius plot draft 1}
\end{figure}

We built our model grid using the stellar and planetary parameters in Table \ref{tab:combined-params}. To emphasize, our modeling framework and conclusions are directly applicable to other temperate sub-Neptunes (equilibrium temperatures 250-300 K) with similar mass–radius characteristics, which we term \textbf{K2-18b analogs}. Figure \ref{fig:mass-radius plot draft 1} showcases the current population of sub-Neptunes with known masses and radii. Potential Hycean-worlds adapted from the list supplied in \citet{Madhusudhan2021} are highlighted within the broader sub-Neptune population, falling within the transition region between super-Earths and sub-Neptunes where both rocky, gas-rich, water-rich compositions remain plausible \citep{Yu2021}.

\begin{table}[!ht]
  \caption{Fixed model planetary and stellar parameters based on K2-18 system}
  \label{tab:combined-params}
  \begin{tabular}{@{}| l |l |l |l |@{}}
    \toprule
    \textbf{Category} & \textbf{Parameter} & \textbf{Value} & \textbf{Source} \\
    \midrule
    \multirow{5}{*}{Planetary}
      & Planet Mass (M$_p$)                                  & $8.63 \pm 1.35\,M_{\oplus}$   & \citet{Benneke2019} \\
      & Planet Radius (R$_p$)                                & $2.610 \pm 0.087\,R_{\oplus}$  & \citet{Benneke2019} \\
      & Equilibrium Temperature (T$_\mathrm{eq}$)               & $	254 \pm 3.9\,\mathrm{K}$         & \citet{Benneke2019} \\
      & Planet Bulk Density ($\rho$$_p$)                               & \(2.67^{+0.52}_{-0.47}\) $\mathrm{g}/\mathrm{cm}^3$
                            & \citet{Benneke2019}\\
      & Surface Gravity ($g_p$)                  & 1234.85 $\mathrm{cm}/\mathrm{s}^2$   & \citet{Benneke2019}          \\
    \addlinespace
    \hline
    \addlinespace
    \multirow{5}{*}{Stellar}
     & Stellar Mass (M$_*$)                       & $	0.4951 \pm 0.0043\,R_{\odot}$         & \citet{Benneke2019}\\
    & Stellar Radius (R$_*$)                         & $	0.4445 \pm 0.0148\,R_{\odot}$         & \citet{Benneke2019}\\
      & Effective Temperature ($T_{\mathrm{eff}}$) 
                                              & $	3457 \pm 39\,\mathrm{K}$          & \citet{Benneke2019} \\
      & Metallicity ($[\mathrm{Fe}/\mathrm{H}]$)& $+0.12 \pm 0.16\,\mathrm{dex}$       & \citet{Sarkis2018} \\
      & Surface Gravity ($\log g_*$)             & $4.786 \pm 0.006$                      & \citet{Stassun2019}  \\
    \bottomrule
  \end{tabular}
\end{table}


\subsection{PICASO: Climate Modeling with Equilibrium Chemistry} \label{subsec:PICASOModeling}

We used {\texttt{PICASO} v3.0} \citep{Batalha2019, Batalha2022, Mukherjee2023} to generate self-consistent pressure-temperature (P-T) profiles and equilibrium chemical abundances for K2-18b-like atmospheres. For each model, we configured \texttt{PICASO} to assume a 1-Dimensional, clear-sky atmosphere under thermochemical equilibrium conditions, neglecting photochemistry and vertical mixing processes that are handled separately by \texttt{VULCAN} (Section \ref{subsec:VULCANmodeling}).

\begin{figure}[ht]
    \centering
    \begin{minipage}{0.48\textwidth}
        \centering
        \includegraphics[width=\linewidth]{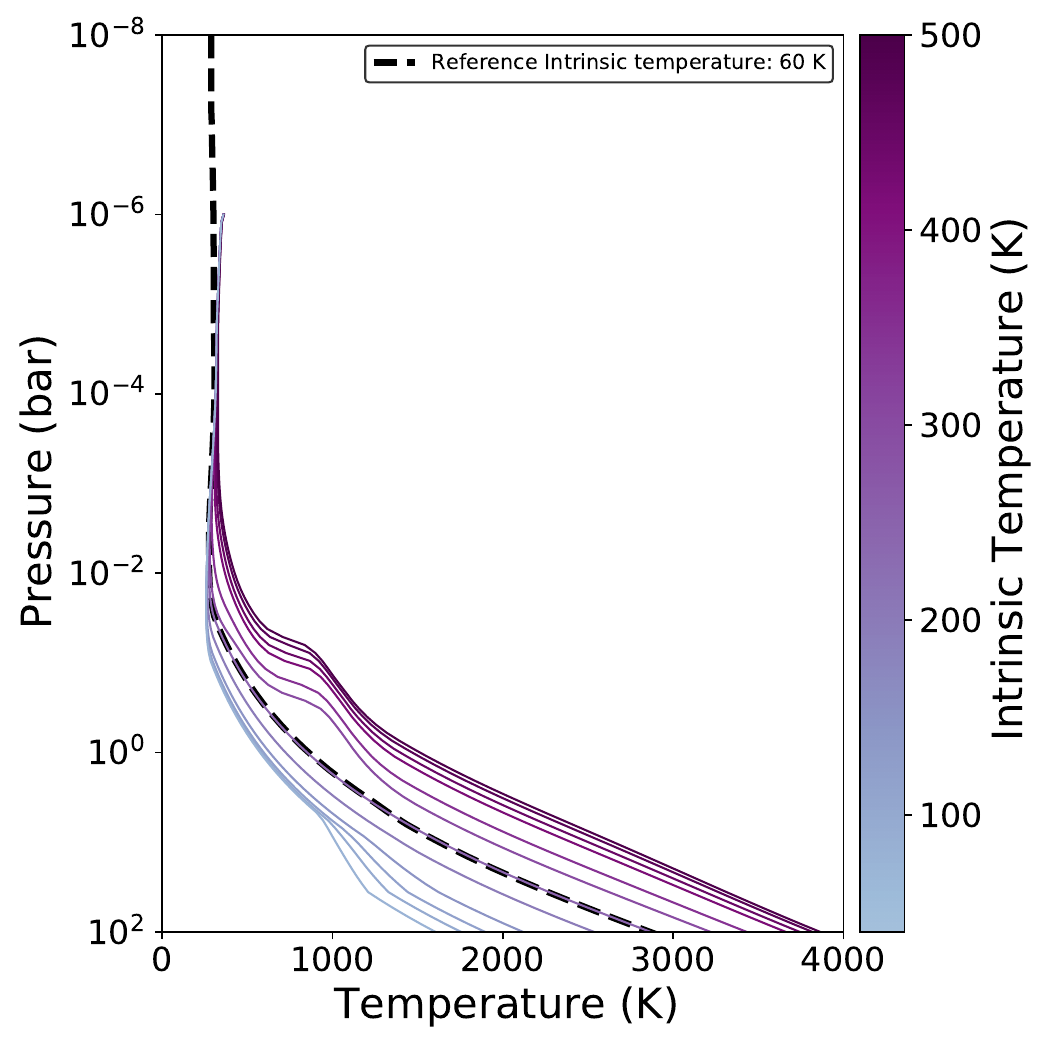}%
    \end{minipage}
    \hfill
    \begin{minipage}{0.48\textwidth}
        \centering
        \includegraphics[width=\linewidth]{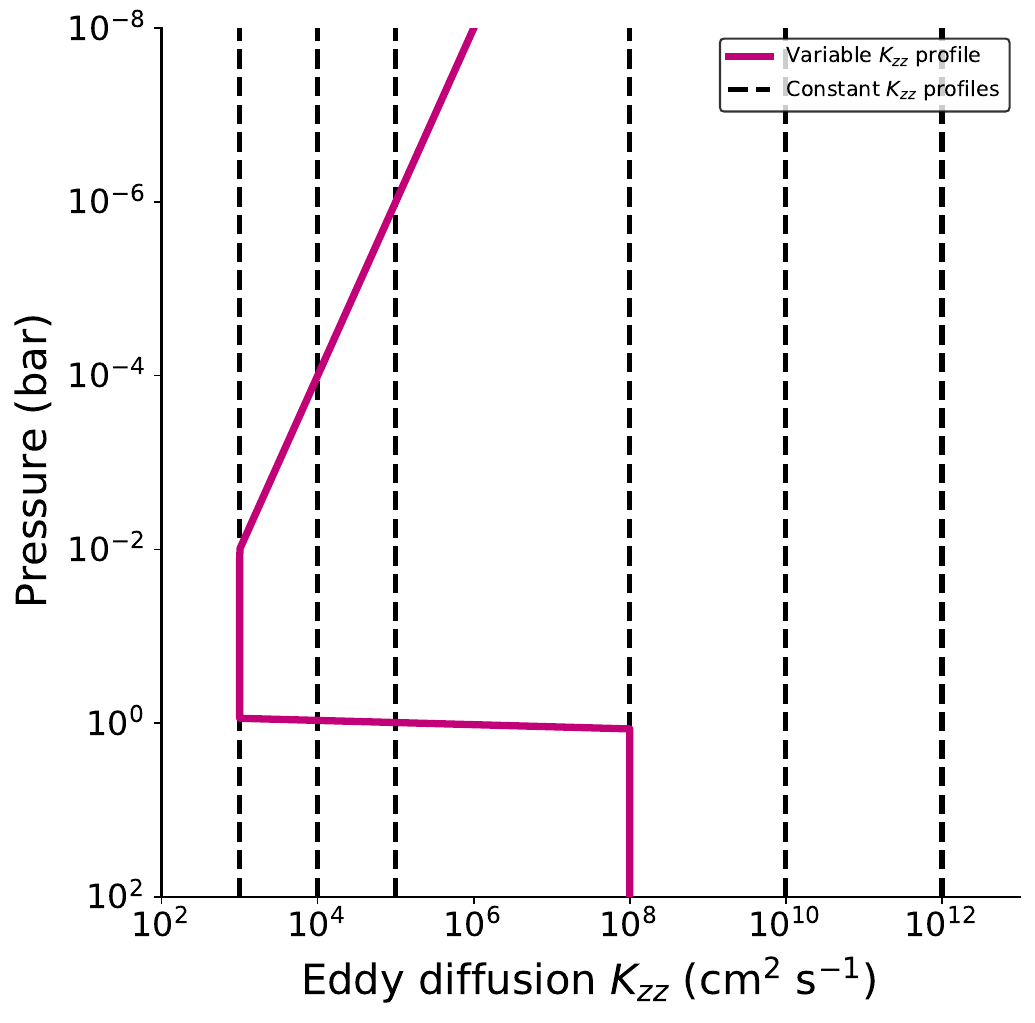}%
    \end{minipage}

    \caption{ \textbf{Left}: Pressure-temperature profiles for a K2-18b-like atmosphere generated by \texttt{PICASO} across varying intrinsic interior temperatures ($T_{\mathrm{int}}$) from $60$ to $450~\mathrm{K}$. The color gradient represents different $T_{\mathrm{int}}$ values, with cooler interiors in blue and hotter interiors in purple. The dashed black line shows the reference case with $T_{\mathrm{int}} = 60~\mathrm{K}$, which has been commonly adopted in previous atmospheric models of K2-18b. The curvature around 0.1 bar shown in Figure \ref{fig:PT-Profile and Kzz Profile} is expected in these PT profiles because that is where the incident stellar irradiation is absorbed by atmospheric gases. \textbf{Right}: Pressure-dependent eddy diffusion coefficient ($K_{\mathrm{zz}}$) profile for a sub-Neptune model of K2-18b with $100\times$ solar metallicity. The purple line shows the model $K_{\mathrm{zz}}$ profile used in this study, which follows a Jupiter-like eddy diffusion profile in the upper atmosphere ($1~\mathrm{bar}$ to $10^{-8}~\mathrm{bar}$) as adopted from \citet{Hu2021, Wogan2024}. The deep atmosphere ($500~\mathrm{bar}$ to $1~\mathrm{bar}$) uses constant $K_{\mathrm{zz}}$ values from $10^{5}$ - $10^{12}$ cm$^2$/s. Black dashed vertical lines indicate the range of constant $K_{\mathrm{zz}}$ values used in our model grid.}
    \label{fig:PT-Profile and Kzz Profile}
\end{figure}

We initialized each \texttt{PICASO} simulation with the planetary and stellar parameters in Table~\ref{tab:combined-params}, assuming a $100\times$ solar metallicity atmosphere and a solar C/O ratio ($\sim$0.5). Given K2-18's M dwarf classification and effective temperature, we employed \texttt{PHOENIX} stellar atmosphere models \citep{Husser2013}. These values are consistent with JWST observations, particularly for K2-18b \citep{Madhusudhan2023, Wogan2024, Jaziri2025}. The atmospheric grid spans 91 pressure levels spaced from $10^{2}$ to $10^{-6}$ bar, with molecular opacities from the correlated-k opacity database of \citet{lupu_2021_7542068}. We used \citet{Guillot2010} pressure-temperature parameterization as the initial P-T profile and set the heat redistribution factor to $r_{\mathrm{facv}} = 0.5$ to represent day-night heat transport (e.g., a planet-wide average P-T profile). The relatively large orbital distance \citep[$a/R_{\star}$ = 48.7;][]{Benneke2019} indicates K2-18b likely experiences efficient day-night heat redistribution through atmospheric circulation, unlike the inefficient transport in tidally locked systems \citep{Leconte2015}.

Our climate and equilibrium chemistry calculations included a H/He-dominated atmosphere, major molecular species (H$_2$O, CO, CO$_2$, CH$_4$, NH$_3$, HCN, C$_2$H$_2$, C$_2$H$_4$), trace gases (NO, OH, CN, PH$_3$), atomic species (Na, K, Fe), ions (H$^+$, H$^-$, e$^-$), and collision-induced absorption from H$_2$-H$_2$ and H$_2$-He pairs \citep{lupu_2021_7542068}. \texttt{PICASO} accounts for refractory species which condense at K2-18b's equilibrium temperature of 250 K and are thus removed from the gas phase in the deep atmosphere.

To explore the impact of interior heating on atmospheric structure, we systematically varied the intrinsic temperature ($T_{\mathrm{int}}$) from 60 K to 450 K at the discrete values 60, 100, 150, 200, 250, 300, 350, 400 and 450 K. The intrinsic temperature parameterizes the internal heat flux from the planet's interior and is related to the effective temperature through
\begin{equation}
T_{\mathrm{eff}}^4 = T_{\mathrm{eq}}^4 + T_{\mathrm{int}}^4,
\end{equation}
where $T_{\mathrm{eq}}$ = 254 K is the equilibrium temperature of K2-18b \citep{Benneke2019} and $T_{\mathrm{eff}}$ is the total effective temperature driving the atmospheric thermal structure \citep{Guillot2010, Fortney2020}. The intrinsic luminosity is given by
\begin{equation}
L_{\mathrm{int}} = 4\pi R_p^2 \sigma T_{\mathrm{int}}^4,
\end{equation}
where $R_p$ is the planetary radius and $\sigma$ is the Stefan-Boltzmann constant. For each $T_{\mathrm{int}}$ value, we computed radiative-convective equilibrium by iteratively adjusting the pressure-temperature profile until energy balance was achieved within a 0.1\% tolerance. We initially attempted to include models with $T_{\mathrm{int}} = 40$~K to extend our parameter space to even cooler interior conditions. However, these models failed to achieve stable radiative convective equilibrium within our convergence tolerance, exhibiting persistent oscillations in the deep atmospheric thermal structure. Since disequilibrium chemistry is fundamentally determined by the deep thermal structure and quench point locations, unconverged pressure-temperature (P-T) profiles would introduce systematic errors in the photospheric chemical abundances that form the basis of our transmission spectra analysis. We therefore excluded $T_{\mathrm{int}} = 40$ K from our final model grid to ensure that all chemical trends reported are based on properly converged atmospheric structures. Figure~\ref{fig:PT-Profile and Kzz Profile} shows the resulting pressure--temperature (P--T) profiles for a planet similar in structure to K2-18b across this range of intrinsic temperatures, illustrating how increased internal heating systematically raises temperatures throughout the deep atmosphere while leaving the upper atmosphere largely unchanged.

For the upper atmosphere, the P-T profile was adapted to become isothermal following the framework in \citet{Wogan2024}. The isothermal temperature was set to the equilibrium temperature of 250 K at the 10 mbar level from the converged radiative convective profile. These outputs served as initial conditions for the subsequent \texttt{VULCAN} disequilibrium chemistry calculations. The curvature around 0.1 bar shown in Figure \ref{fig:PT-Profile and Kzz Profile} is expected in these PT profiles because that is where the incident stellar irradiation is absorbed by atmospheric gases.


\subsection{VULCAN: Disequilibrium and Photochemistry} \label{subsec:VULCANmodeling}

We used \texttt{VULCAN} \citep{Tsai2021} to compute disequilibrium chemical abundances, taking the P-T profiles generated by \texttt{PICASO} (see Section \ref{subsec:PICASOModeling}) as input and evolving the atmospheric chemistry beyond thermochemical equilibrium. For each \texttt{PICASO} model, we configured \texttt{VULCAN} to include photochemistry, vertical mixing, and condensation processes for H$_2$O that are essential to simulate the atmospheric conditions of planets like K2-18b. 

Our atmosphere was configured as H/He-dominated to reflect a gas-rich envelope. We again used 91 vertical layers, spanning a wide pressure range from a bottom pressure \(P_{b} = 10^{3}\,\mathrm{bar}\) to a top pressure \(P_{t} = 10^{-8}\,\mathrm{bar}\), to capture the full vertical extent of the observable atmosphere similar to Section \ref{subsec:PICASOModeling}. To account for vertical transport, we enabled both eddy and molecular diffusion. We specified a range of constant eddy diffusion coefficients for the Constant \(K_{zz}\) Profile from $10^{5}$ - $10^{12}$ cm$^2$/s to systematically explore the effects of vertical mixing on the atmospheric chemistry and followed the $K_{\mathrm{zz}}$ profile from \citep{Wogan2024} for the Variable \(K_{zz}\) Profile. Large scale vertical advection was disabled to isolate the influence of other transport mechanisms.

We implemented the NCHO chemical network containing around 100 NCHO species with thermodynamically consistent reversible reactions. We adopted a customized set of elemental abundances, specifically assuming $100\times$ solar metallicity and a solar C/O ratio. This choice was motivated by prior modeling from \citet{Wogan2024}, which showed that volatile enriched enriched atmospheres can reproduce the inferred composition of K2-18b, including strong CO$_2$ and CH$_4$ features and the absence of NH$_3$. We also included sulfur chemistry in our initial configuration, but did not include the full SNCHO network in our model grids for computational simplicity. While sulfur photochemistry can produce various sulfur-bearing species under disequilibrium conditions, we focused on the NCHO network to isolate the effects of intrinsic temperature and vertical mixing on the primary carbon, nitrogen, and oxygen chemistry that drives our spectral diagnostics. We verified this approach by running a benchmark SNCHO test, which showed an difference of 1.8 ppm compared to our NCHO models.

To verify that the predicted composition remained effectively unchanged when sulfur chemistry is included, we ran an SNCHO test for a benchmark case adopting the variable $K_{\mathrm{zz}}$ profile and $T_{\mathrm{int}}=60$~K. A comparison of the resulting transmission spectra over 0.6--5.3~$\mu$m (Section~\ref{subsec: Transmission Spectrum Calculation}) yields wavelength-averaged differences of $1.2 \pm 0.5$~ppm between the NCHO and SNCHO models (Figure \ref{fig:stacked_two}). The largest difference occurred in the CO$_2$ absorption region near 4.3~$\mu$m ($\leq 7$~ppm) and in the CO feature at 4.6~$\mu$m ($4$~ppm), while CH$_4$ diagnostic features at 2.3 and 3.3~$\mu$m show $<3$~ppm sensitivity and H$_2$O features remain essentially unchanged ($<2$~ppm). The overall difference of 1.8~ppm confirms that the inclusion of sulfur chemistry does not alter the key molecular abundances or introduce systematic biases in our primary diagnostic species (CH$_4$, CO$_2$, H$_2$O, NH$_3$). We therefore adopt the simpler NCHO chemical network for the remainder of this work.

M-dwarf stars are particularly effective drivers of atmospheric photochemistry due to their enhanced UV emission relative to their bolometric luminosity, making the stars critical for understanding molecular dissociation and disequilibrium chemistry in sub-Neptune atmospheres \citep{Hu2012, France2013}. K2-18, classified as an M3-dwarf star with an effective temperature of 3457 $\pm$ 39 K (Table \ref{tab:combined-params}), provides an ideal case study for these photochemical processes \citep{Benneke2019}. Since UV measurements of K2-18 are not available, we adopted the GJ-176 stellar flux from \citet{France2013}, modified by \citet{Wogan2024} as a proxy for K2-18 stellar flux. However, K2-18 appears to be more active than typical M dwarfs of its age, with multiple spot crossing events observed in Kepler data indicating significant stellar activity and variability \citep{Barclay2021}. While we assume the largely inactive GJ-176 UV spectrum for this analysis, the enhanced activity of K2-18 could potentially increase photochemical reaction rates and molecular dissociation beyond our current estimates. We used a standard two-stream Eddington coefficient of 0.5 to model the balance of upward and downward UV scattering in \texttt{VULCAN}. The model was configured with the planet's known physical parameters (Table \ref{tab:combined-params})

K2-18b's equilibrium temperature of 254 $\pm$ 3.9 K (assuming albedo = 0.3; Table \ref{tab:combined-params}) indicates the potential for water condensation in the lower atmosphere. We enabled the condensation mode in \texttt{VULCAN} for H$_2$O only. We disabled gravitational settling of condensate particles to prevent excessive removal of water vapor from the upper atmosphere, which would unrealistically deplete stratospheric (H\(_2\)O). This approach is consistent with recent atmospheric modeling practices for K2-18b \citep{Wogan2024} and similar sub-Neptune exoplanets \citep{Tsai2024}, where careful treatment of condensation and transport processes is essential for accurate photochemical modeling

\subsection{Generating Transmission Spectrum}\label{subsec: Transmission Spectrum Calculation}

Transmission spectra were computed using {\texttt{PICASO} v3.0} by inputting the atmospheric composition profiles generated by \texttt{VULCAN}. For each atmospheric model, we extracted the pressure-temperature profile and molecular mixing ratios from \texttt{VULCAN} output and fed them into {\texttt{PICASO}}'s radiative transfer calculations. We configured \texttt{PICASO} with K2-18b's measured planetary parameters and stellar properties from Table \ref{tab:combined-params}. Opacity calculations used a comprehensive molecular line database spanning 0.6–5.3 $\mu\mathrm{m}$ wavelengths at R = 10,000 spectral resolution \citep{Marley2021}. While the $R = 60{,}000$ database includes additional molecular species, we verified that using the higher-resolution database produced no significant spectral differences across our wavelength range for the atmospheric conditions and molecular abundances in our models. The $R = 10{,}000$ database includes the primary opacity contributors for our analysis: CH$_4$, CO, CO$_2$, H$_2$O, NH$_3$, H$_2$S, and alkali metals. The transmission spectrum calculation employed \texttt{PICASO}'s 1-D radiative transfer assuming a clear-sky atmosphere. Final spectra were binned to R = 100 resolution to match the JWST-NIRSpec/PRISM observations and facilitate comparison with published K2-18b spectra. Computing the transmission spectrum with coupled \texttt{VULCAN}-\texttt{PICASO} approach ensures self-consistent atmospheric composition and radiative transfer, providing realistic transmission spectra that account for both equilibrium and disequilibrium chemistry across our full parameter grid.

\subsection{Benchmark Model} \label{subsec:Benchmark Model}

As we combined the radiative-convective and equilibrium chemistry code \texttt{PICASO v3.0} with the disequilibrium chemistry and photochemistry code \texttt{VULCAN} to create model grids for K2-18b analogs, our approach differs from \citet{Wogan2024}, who employed \texttt{PICASO} coupled with the \texttt{PHOTOCHEM} code \citep{Wogan2025}. To validate our modeling framework, we benchmark our results against the gas-rich sub-Neptune model from \citet{Wogan2024}. While \citet{Wogan2025} demonstrated that \texttt{PHOTOCHEM} and \texttt{VULCAN} are highly consistent and produce nearly identical atmospheric compositions for the same inputs, our use of \texttt{VULCAN} provides another comprehensive photochemical and transport-driven disequilibrium network. The differences between packages arise mainly from their independent chemical kinetics databases, making both approaches complementary. Our \texttt{PICASO}-\texttt{VULCAN} framework enabled us to conduct a robust exploration of how interior heating and mixing processes shape observable spectral signatures in K2-18b and its analogs.

Using identical P--T profiles, variable variable ($K_{\mathrm{zz}}$)profiles, stellar profiles, atmospheric metallicity, C/O ratio, elemental abundances, planetary mass and radius, orbital parameters, and boundary conditions from \citet{Wogan2024}, our \texttt{VULCAN} simulation of K2-18b shows our predicted composition (Figure \ref{fig:stacked_two}) matches across 0.5--5 $\mu\mathrm{m}$. The most notable difference occurs around 2.5--3.0 $\mu\mathrm{m}$, where our models exhibit 15.7 ppm enhanced absorption. The rest of the spectrum shows excellent agreement of less than 9 ppm differences. The difference appears to be how condensation is handled between the two codes. We attempted more aggressive condensation treatment to force that wavelength range to match the \citet{Wogan2024} spectrum, but this led to significantly depleted CO and CO$_2$ abundances, likely due to the removal of oxygen through enhanced water condensation. Outside of this wavelength range, we are in agreement with the \citet{Wogan2024} results. Other minor discrepancies may also be due to variations in reaction rate coefficients, molecular opacity databases, condensation treatments, or the specific implementation of vertical mixing and photochemical coupling between the two codes. The small magnitude of these differences and the good agreement between the two modeling systems confirms that our \texttt{PICASO}--\texttt{VULCAN} combined approach can be trusted and that our systematic exploration of intrinsic temperature and vertical mixing parameter space provides robust conclusions about atmospheric composition trends, with chemical network and condensation treatment uncertainties secondary to the dominant effects of thermal and dynamical parameters we investigate.

\section{Results} \label{sec:Results}

We present molecular abundances from our grid of atmospheric models for K2-18b spanning intrinsic temperatures ($T_{\mathrm{int}}$) from $60$ to $450\ \mathrm{K}$ and eddy diffusion coefficients ($K_{\mathrm{zz}}$) from $10^{5}$ - $10^{12}$ cm$^2$/s, along with the variable $K_{\mathrm{zz}}$ profile from \citet{Hu2021, Wogan2024} in Figure \ref{fig:PT-Profile and Kzz Profile}. The parameter space encompasses a full range of plausible thermal and mixing states for temperate sub-Neptune atmospheres. Our analysis focuses on three complementary aspects of atmospheric structure and composition: mean molecular abundances in the photospheric region of pressures $10^{-3}\text{-}10^{-4}$ bars (Section \ref{sec:Tint+Kzz Effects}) probed by transmission spectroscopy, vertical abundance profiles revealing quenching and photochemical processes (Section \ref{sec:Quenching}), and transmission spectra showing observable signatures across $1$–$5\ \mu\mathrm{m}$ (Section \ref{sec: spectra}). All abundance values discussed below are reported as $\log_{10}$ volume mixing ratio unless otherwise noted.

\subsection{Effect of Intrinsic Temperature and Vertical Mixing on Molecular Abundances} \label{sec:Tint+Kzz Effects}

\begin{figure}[!h]
  \centering

  \begin{subfigure}[t]{0.3\textwidth}
    \includegraphics[width=1.18\linewidth]{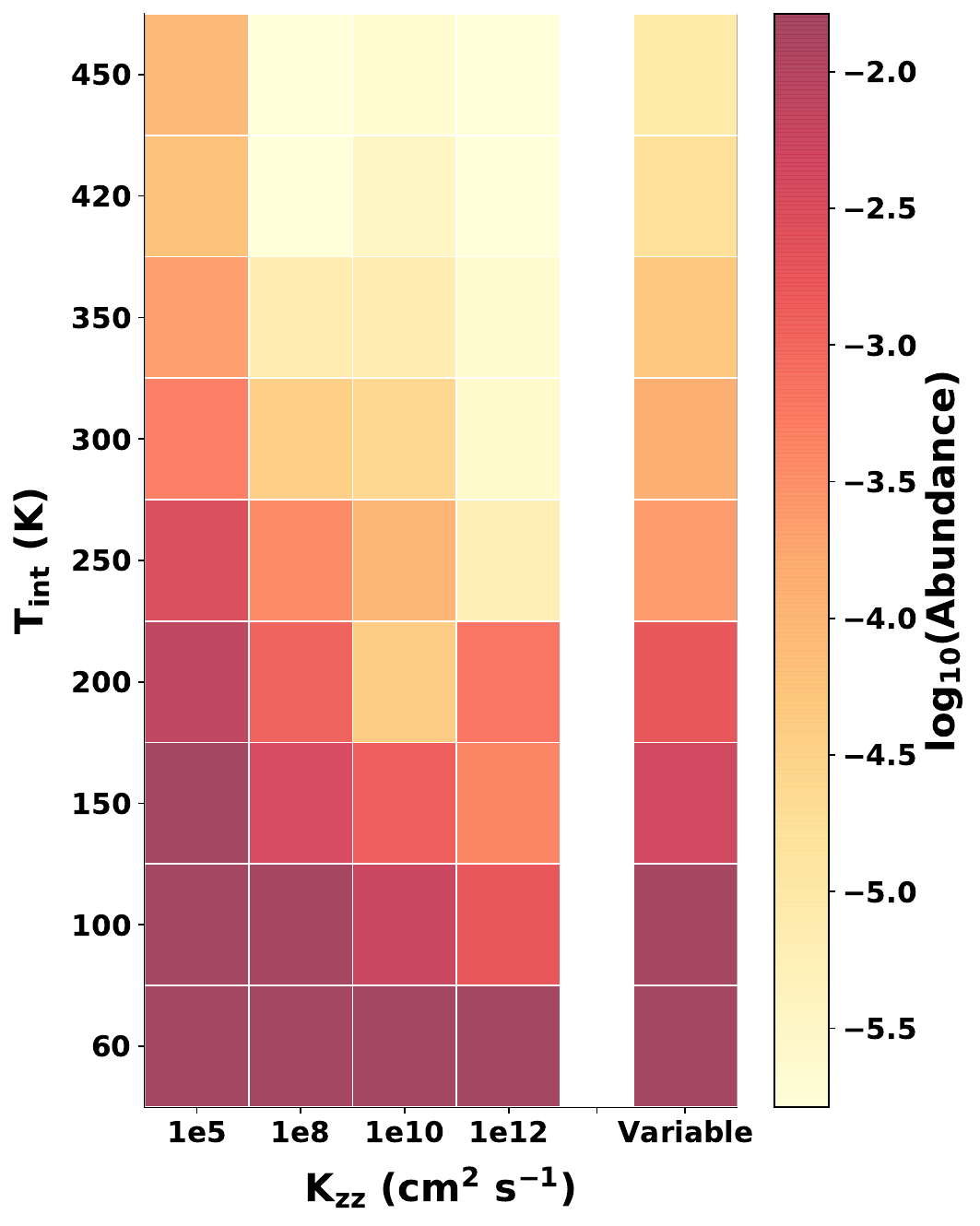}
    \caption{Methane (CH\(_4\))}
    \label{fig:ch4}
  \end{subfigure}\hfill
  \begin{subfigure}[t]{0.3\textwidth}
    \includegraphics[width=1.18\linewidth]{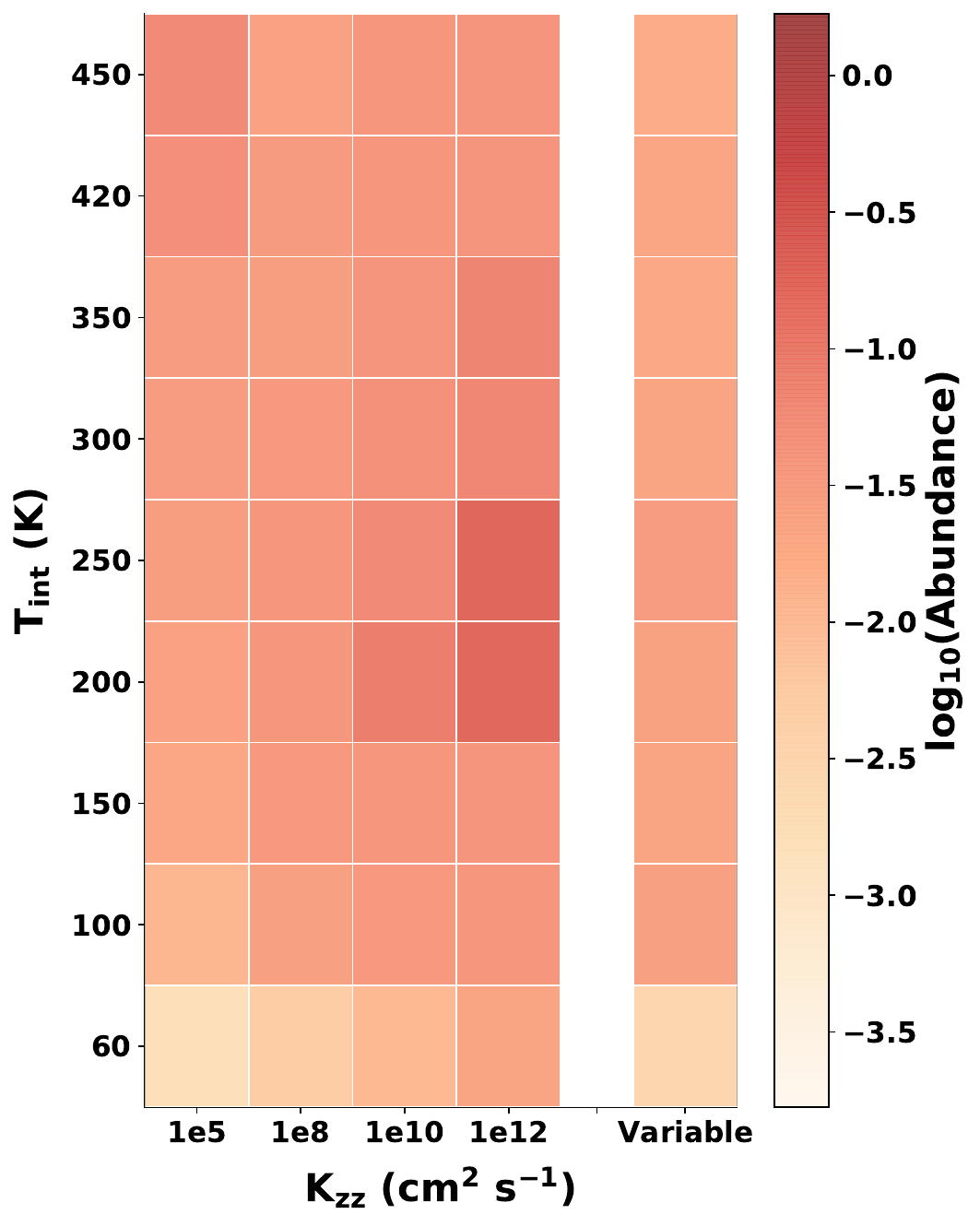}
    \caption{Carbon monoxide (CO)}
    \label{fig:co}
  \end{subfigure}\hfill
  \begin{subfigure}[t]{0.3\textwidth}
    \includegraphics[width=1.18\linewidth]{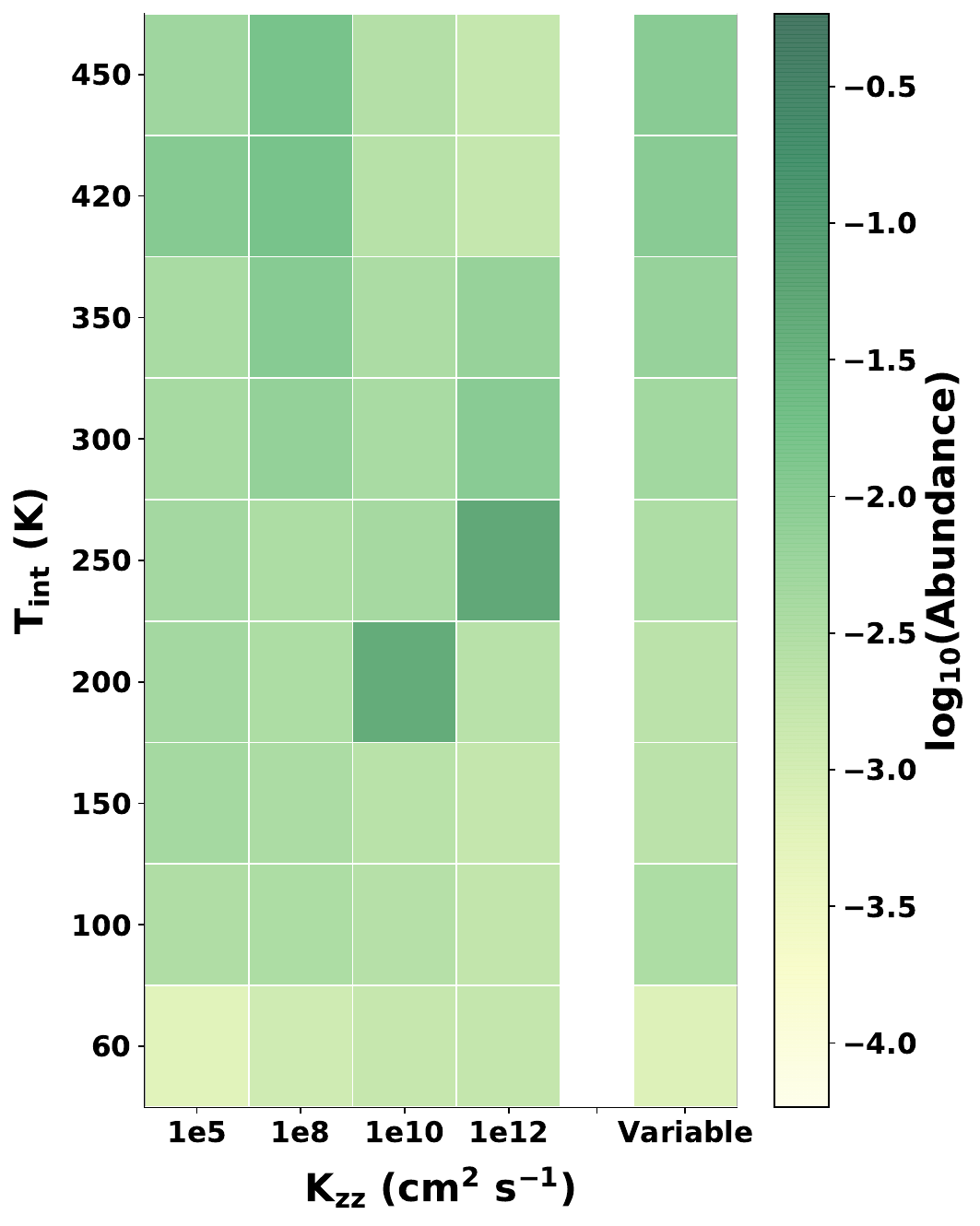}
    \caption{Carbon dioxide (CO\(_2\))}
    \label{fig:co2}
  \end{subfigure}

  \par\bigskip 

  \begin{subfigure}[t]{0.3\textwidth}
    \includegraphics[width=1.18\linewidth]{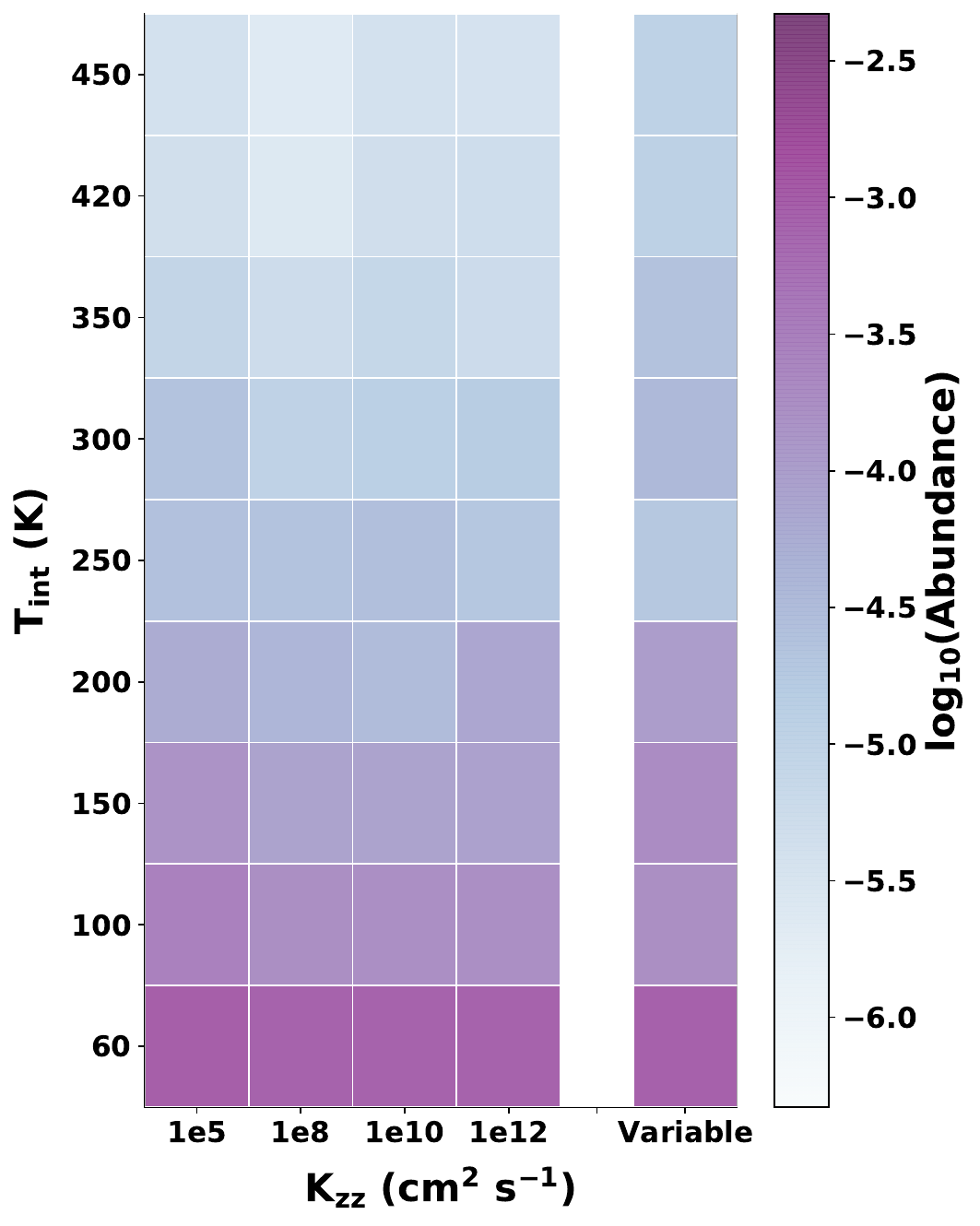}
    \caption{Ammonia (NH\(_3\))}
    \label{fig:nh3}
  \end{subfigure}\hfill
  \begin{subfigure}[t]{0.3\textwidth}
    \includegraphics[width=1.18\linewidth]{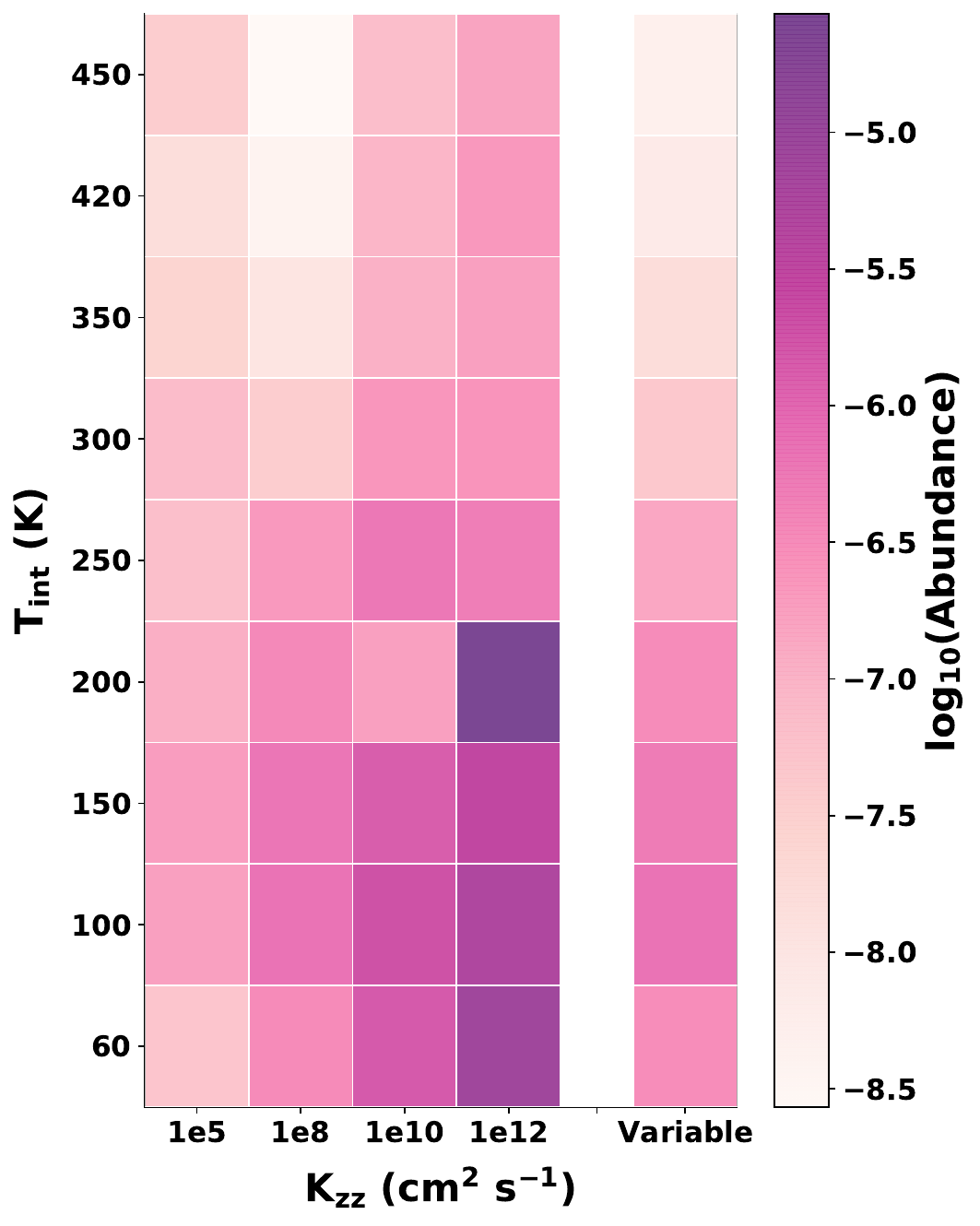}
    \caption{Hydrogen Cyanide (HCN)}
    \label{fig:hcn}
  \end{subfigure}\hfill
  \begin{subfigure}[t]{0.3\textwidth}
    \includegraphics[width=1.18\linewidth]{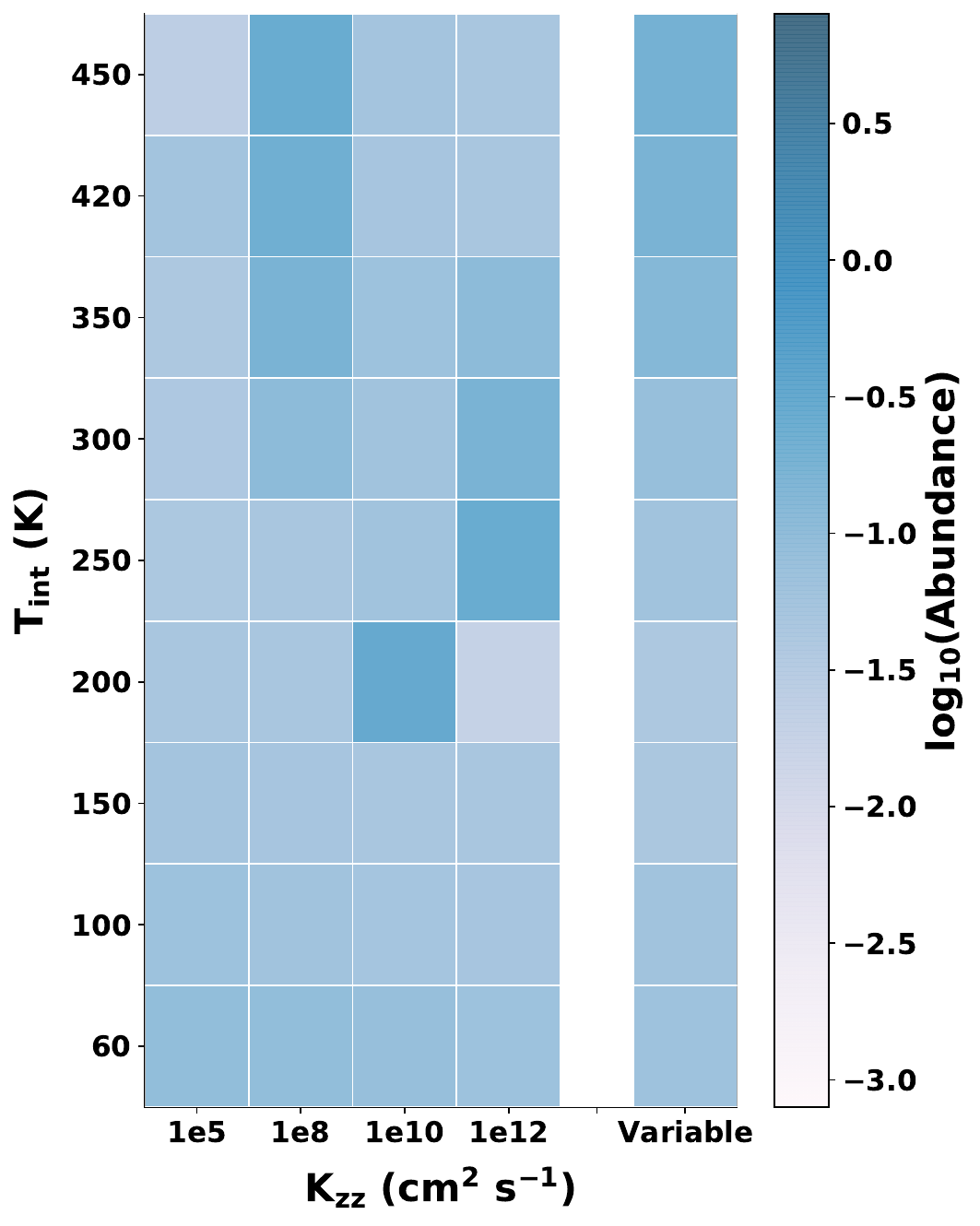}
    \caption{Water (H\(_2\)O)}
    \label{fig:h2o}
  \end{subfigure}

  \caption{Mean $\log_{10}$ abundance averaged over $10^{-3}$ -- $10^{-4}$ bar as a function of intrinsic temperature ($60$ -- $450$ K, y\mbox{-}axis) and vertical eddy diffusion coefficient $K_{\mathrm{zz}}$ ($10^{5}$ - $10^{12}\ \mathrm{cm}^{2}\ \mathrm{s}^{-1}$, x\mbox{-}axis) for key atmospheric molecules: (a) methane (CH$_4$): yellow to orange, (b) carbon monoxide (CO): red dark to red, (c) carbon dioxide (CO$_2$): light green to dark green, (d) ammonia (NH$_3$): light blue to purple), (e) hydrogen cyanide (HCN): pink to magenta), and (f) water (H$_2$O): light blue to dark blue. More saturated colors (Darker) indicate higher abundances, while lighter colors indicate depletion. The column labeled \textbf{Variable} represents modeled grid using $K_{\mathrm{zz}}$ profile shown in Figure \ref{fig:PT-Profile and Kzz Profile} as adopted from \citet{Hu2021} and \citet{Wogan2024}.}
  \label{fig:abundances-six}
\end{figure}

Figure \ref{fig:abundances-six} displays mean abundances of six key molecules: CH$_4$, CO, CO$_2$, NH$_3$, HCN, and H$_2$O averaged between pressures $10^{-3}$ and $10^{-4}$ bar. These pressure levels correspond to the photospheric region probed by transmission spectroscopy \citep{Madhusudhan2023}. The abundances at these altitudes reflect the complex interplay between deep atmospheric chemistry, vertical transport (quenching), and upper atmospheric photochemistry \citep{Visscher2010, Hu2021, Tsai2021}. 

Our systematic exploration reveals distinct molecular behavior across three thermal regimes: (1) Cool temperature models ($T_{\mathrm{int}} \leq 150$ K), (2) Intermediate temperature models ($150$ -- $350\ \mathrm{K}$), and (3) High temperature models ($T_{\mathrm{int}} \geq 350$ K). These regimes exhibit fundamentally different chemical characteristics: temperature dominated species insensitive to mixing (CH$_4$, NH$_3$), moderately sensitive species showing coupled thermal dynamical effects (CO, H$_2$O, CO$_2$), and extreme disequilibrium tracers whose abundances vary by orders of magnitude with mixing strength (HCN). We organize our discussion using these temperature regime classifications throughout this section unless specific temperature values are noted. 

When comparing the variable $K_{\mathrm{zz}}$ profile from \citet{Wogan2024} (Figure~\ref{fig:PT-Profile and Kzz Profile}) to constant $K_{\mathrm{zz}}$ profiles, different molecular species exhibit distinct sensitivity patterns that reveal the underlying atmospheric chemistry controls. CH$_4$ and NH$_3$ exhibit no difference between variable and constant $K_{\mathrm{zz}}$ cases, confirming thermochemical equilibrium dominance over vertical transport effects for these temperature-sensitive species. CO and H$_2$O (Figures \ref{fig:co} and \ref{fig:h2o}) show differences, with the variable profile tracking intermediate constant $K_{\mathrm{zz}}$ cases ($10^{8}$ - $10^{10}$ cm$^2$/s) and differing by $<0.5$~dex across most temperatures. CO$_2$ (Figure \ref{fig:co2}) shows enhanced abundances of $10^{-2}$ to $10^{-2.5}$ in intermediate temperature regimes with the variable $0.5$--$1.0$~dex increase over most constant $K_{\mathrm{zz}}$ cases. Most notably, HCN (Figure \ref{fig:hcn}) exhibits the most dramatic behavioral change, shifting from primary $K_{\mathrm{zz}}$ sensitivity, with up to six orders of magnitude variation, in constant profiles to strongly temperature-dependent behavior with the variable $K_{\mathrm{zz}}$ profile. This indicates that altitude-dependent mixing creates asymmetric disequilibrium effects across thermal regimes \citep{Visscher2010, Tsai2021}.

\subsubsection{CH$_4$, CO, CO$_2$}
\label{sec:carbon_species}

In cool intrinsic-temperature regimes, thermochemical equilibrium strongly favors CH$_4$ production through
\begin{equation}
\mathrm{CO} + 3\,\mathrm{H}_2 \rightarrow \mathrm{CH}_4 + \mathrm{H}_2\mathrm{O}\,,
\label{eq:3}
\end{equation}
such that CH$_4$ reaches abundances of $10^{-2}$ to $10^{-3}$ and exhibits strong temperature dependence but weak $K_{\mathrm{zz}}$ sensitivity, indicating thermochemical-equilibrium control \citep{Visscher2006, Moses2011}. As temperature increases to intermediate regimes ($200$--$350$~K), CH$_4$ abundances decrease to $10^{-3}$ to $10^{-4}$, while at high temperatures ($>350$~K) severe depletion occurs, with abundances dropping to $10^{-5}$ to $10^{-6}$ as the equilibrium reverses to favor CO production.

CO exhibits weaker temperature sensitivity, with abundances ranging from $10^{-0.5}$ to $10^{-1}$ across most of parameter space, and shows systematic enhancement at higher intrinsic temperatures as it becomes the thermodynamically favored carbon carrier \citep{Visscher2006}. Unlike CH$_4$, CO is photochemically stable due to its strong C$\equiv$O triple bond, making it largely immune to stellar-UV destruction.

CO$_2$ demonstrates the most complex behavior, reaching peak abundances of $10^{-2}$ to $10^{-2.5}$ in intermediate-temperature models where it serves as a diagnostic for moderately heated interior conditions. CO$_2$ becomes depleted in both cool and hot regimes, but shows strong $K_{\mathrm{zz}}$ dependence under intermediate conditions.

The dominance of interior thermal processes over photochemistry becomes clear when examining carbon speciation. While upper-atmosphere photochemical pathways produce negligible CO and CO$_2$, deep hot interiors dramatically enhance these species through efficient vertical transport of thermochemically equilibrated gas from depth \citep{Moses2011, Visscher2011}. This deep quenching process simultaneously suppresses CH$_4$ before it reaches the photochemical zone, creating a fundamental asymmetry: CH$_4$ can be destroyed by both thermal equilibrium and photochemistry, whereas CO and CO$_2$ are produced thermally and remain photochemically stable.

\subsubsection{NH$_3$}
\label{sec:nitrogen_nh3}

NH$_3$ behavior reflects the complex interplay between thermochemical equilibrium and photochemical processes in nitrogen chemistry \citep{Heng2016}. In cool temperature regimes, thermochemical equilibrium favors NH$_3$ production through
\begin{equation}
\mathrm{N}_2 + 3\,\mathrm{H}_2 \rightarrow 2\,\mathrm{NH}_3\,,
\label{eq:4}
\end{equation}
producing high abundances of $10^{-2.5}$ to $10^{-3}$ (Figure~\ref{fig:nh3}). However, NH$_3$ chemistry is fundamentally more complex than carbon chemistry due to the chemical inertness of N$_2$. NH$_3$ is thermochemically favored under cool temperatures and high pressures, but the equilibrium with N$_2$ \,($2\,\mathrm{NH}_3 \rightleftharpoons \mathrm{N}_2 + 3\,\mathrm{H}_2$)\, becomes increasingly unfavorable as temperature rises, since N$_2$ is chemically inert and the reaction kinetics are sluggish.

This thermal sensitivity is compounded by photochemical fragility. Unlike the carbon-bearing species CO and CO$_2$, which are photochemically stable, NH$_3$ is readily destroyed by stellar UV radiation \citep{Yu2021, Hu2021, Tsai2021}. The degree of photochemical depletion depends critically on stellar type, as M dwarf stars provide the near-UV irradiation needed for NH$_3$ photodissociation \citep{Yu2021}.

However, our models using the K2-18 stellar flux from \citet{Wogan2024} show NH$_3$ abundances remaining relatively constant across our grid, with high intrinsic temperature thermal depletion to $10^{-5.5}$ being the dominant effect rather than photochemical destruction, despite the known stellar activity of this M dwarf \citep{Barclay2021}. This thermal control over NH$_3$ abundances, combined with its weak $K_{\mathrm{zz}}$ sensitivity, indicates that interior temperature rather than atmospheric mixing or photochemistry primarily determines observable nitrogen speciation in our K2-18b analog models.

\subsubsection{H$_2$O}\label{subsubsec: h2o}

H$_2$O exhibits minimal sensitivity to both intrinsic temperature and vertical mixing strength (Figure~\ref{fig:h2o}). H$_2$O maintains stable abundances across the entire parameter space, ranging from approximately $10^{-0.5}$ to $10^{-1.5}$, showing little systematic variation (Appendix Table~\ref{tab:Abundance_trends}). This behavior reflects two key factors: (1) H$_2$O's thermodynamic stability in hydrogen-dominated atmospheres, where it remains the dominant oxygen-bearing species across a wide range of conditions \citep{Visscher2006}, and (2) the fact that our atmospheric temperature profiles remain above the H$_2$O condensation curve throughout most of the observable atmosphere. Water condensation occurs at approximately 270~K at 1~bar, but our pressure--temperature profiles show temperatures exceeding this threshold at 1~bar for most intrinsic-temperature models, preventing significant condensation depletion in the photospheric region ($10^{-3}$--$10^{-4}$~bar) probed by transmission spectroscopy.

The uniform abundance distribution across our intrinsic temperature range (60--450 K) and vertical mixing strength range ($10^{5}$ - $10^{12}$ cm$^2$/s) indicates that H$_2$O abundances are controlled primarily by the bulk atmospheric metallicity and C/O ratio rather than by local chemical equilibrium or transport processes. Additionally, part of this muted response reflects the fundamental control of the C/O ratio on water abundance. With our assumed solar C/O ratio (0.5), excess oxygen beyond that consumed by CO formation is available to form H$_2$O, maintaining relatively stable abundances across thermal conditions. In high C/O ratio atmospheres (C/O $>$ 1), carbon would consume nearly all available oxygen to form CO, leaving water virtually absent regardless of metallicity or temperature \citep{Madhusudhan2012, Moses2011}. This makes H$_2$O a valuable diagnostic of atmospheric C/O ratio, with its abundance primarily determined by the initial elemental composition rather than local thermochemical processes \citep{Oberg2011, Madhusudhan2012}.

Even at the highest intrinsic temperatures, where other molecules like CH$_4$ and NH$_3$ show dramatic depletion, H$_2$O remains comparatively unaffected due to its high thermodynamic stability and its role as a primary oxygen reservoir under these conditions. This insensitivity to both thermal and mixing parameters makes H$_2$O a poor diagnostic for constraining interior thermal structure or atmospheric dynamics, but an excellent tracer of bulk atmospheric composition and metallicity in sub-Neptune atmospheres.

\subsubsection{HCN}\label{subsubsec: hcn}

Hydrogen cyanide (HCN) exhibits the most complex atmospheric behavior, with extreme sensitivity to both $T_{\mathrm{int}}$ and $K_{\mathrm{zz}}$ (Figure~\ref{fig:hcn}). Our analysis reveals that photochemistry strongly couples NH$_3$ and HCN abundances, with HCN becoming the primary photochemical product of NH$_3$ in CH$_4$-rich environments \citep{Moses2011, Hu2021, Tsai2023}. 

HCN abundances displays variability across our parameter grid. At high temperatures with low $10^{5}$ cm$^2$/s, HCN reaches peak abundances of approximately $10^{-5}$ to $10^{-7}$. However, the $K_{\mathrm{zz}}$ sensitivity is extreme: abundances vary by up to $6$ orders of magnitude as $K_{\mathrm{zz}}$ increases from $10^{5}$ - $10^{12}$ cm$^2$/s. For $K_{zz}\gtrsim$ $10^{10}$ cm$^2$/s, HCN abundances plummet to the lowest values in our grid, while at low temperatures, HCN becomes depleted regardless of mixing strength (Appendix Table~\ref{tab:Abundance_trends}).

This extraordinary sensitivity reflects competing formation and destruction processes operating across different atmospheric regions. HCN forms through coupled NH$_3$ + CH$_4$ photochemistry in the upper atmosphere ($10^{-4}$--$10^{-2}$~bar), where stellar UV drives the reaction sequence:
\begin{align}
\mathrm{NH_3} + h\nu &\rightarrow \mathrm{NH_2} \rightarrow \mathrm{NH} \rightarrow \mathrm{N} \\
\mathrm{CH_4} + h\nu &\rightarrow \mathrm{CH_3} \\
\mathrm{N} + \mathrm{CH_3} &\rightarrow \mathrm{H_2CN} \rightarrow \mathrm{HCN}.
\end{align}
The availability of atomic nitrogen from NH$_3$ photodissociation proves critical for HCN formation, making this species directly dependent on both parent molecules \citep{Zahnle2009, Moses2011, Tsai2021}.

However, intrinsic temperature exerts indirect control by governing the initial supply of NH$_3$ and CH$_4$ from the deep atmosphere. Additionally, while HCN can form efficiently in hot interior layers, it faces rapid destruction in cooler upper regions through UV photolysis and oxidation by O and OH radicals \citep{Rimmer2020}. This creates a fundamental competition: weak mixing (low $K_{\mathrm{zz}}$) preserves HCN formed at depth, while strong mixing rapidly transports HCN to destruction-dominated regions \citep{Zahnle2009, Moses2011, Hu2021}. The extreme $K_{\mathrm{zz}}$ sensitivity therefore reflects this balance between transport-controlled quenching in the deep interior and photochemical processing in the observable atmosphere, making HCN a diagnostic of both thermal structure and atmospheric dynamics.

However, intrinsic temperature exerts indirect control by governing the initial supply of NH$_3$ and CH$_4$ from the deep atmosphere that serve as HCN precursors. HCN formation occurs primarily in the upper atmosphere ($10^{-4}$ to $10^{-2}$~bar) through photochemical pathways requiring UV-dissociated NH$_3$ and CH$_4$, rather than through thermal processes in the deep interior where thermochemical equilibrium favors N$_2$ and CH$_4$. Once formed photochemically, HCN faces rapid destruction in the upper atmosphere through continued UV photolysis and oxidation by O and OH radicals \citep{Rimmer2020}. This creates a fundamental competition: weak mixing (low $K_{\mathrm{zz}}$) allows HCN to accumulate in its formation region, while strong mixing rapidly transports HCN to destruction-dominated regions or dilutes it with HCN-poor gas from below \citep{Zahnle2009, Moses2011, Hu2021}. The extreme $K_{\mathrm{zz}}$ sensitivity therefore reflects this balance between photochemical production and destruction in the observable atmosphere, making HCN a diagnostic of both photochemical activity and atmospheric dynamics.

\subsection{Vertical Abundance Profiles and Atmospheric Structure}\label{sec:Quenching}

Quenching occurs when vertical mixing timescales become comparable to chemical equilibration timescales, causing molecular abundances to freeze out and deviate from local thermochemical equilibrium \citep{Visscher2010, Moses2011}. The quench pressure, the atmospheric level where this transition from equilibrium to transport control occurs, varies systematically with $K_{\mathrm{zz}}$, $T_{\mathrm{int}}$, and molecular species \citep{Visscher2010, Moses2011}.

\captionsetup[subfigure]{skip=-1pt}
\captionsetup[figure]{skip=4pt}
\begin{figure}[!h]
  \centering

  \begin{subfigure}[t]{0.48\textwidth}
    \centering
    \includegraphics[width=\linewidth]{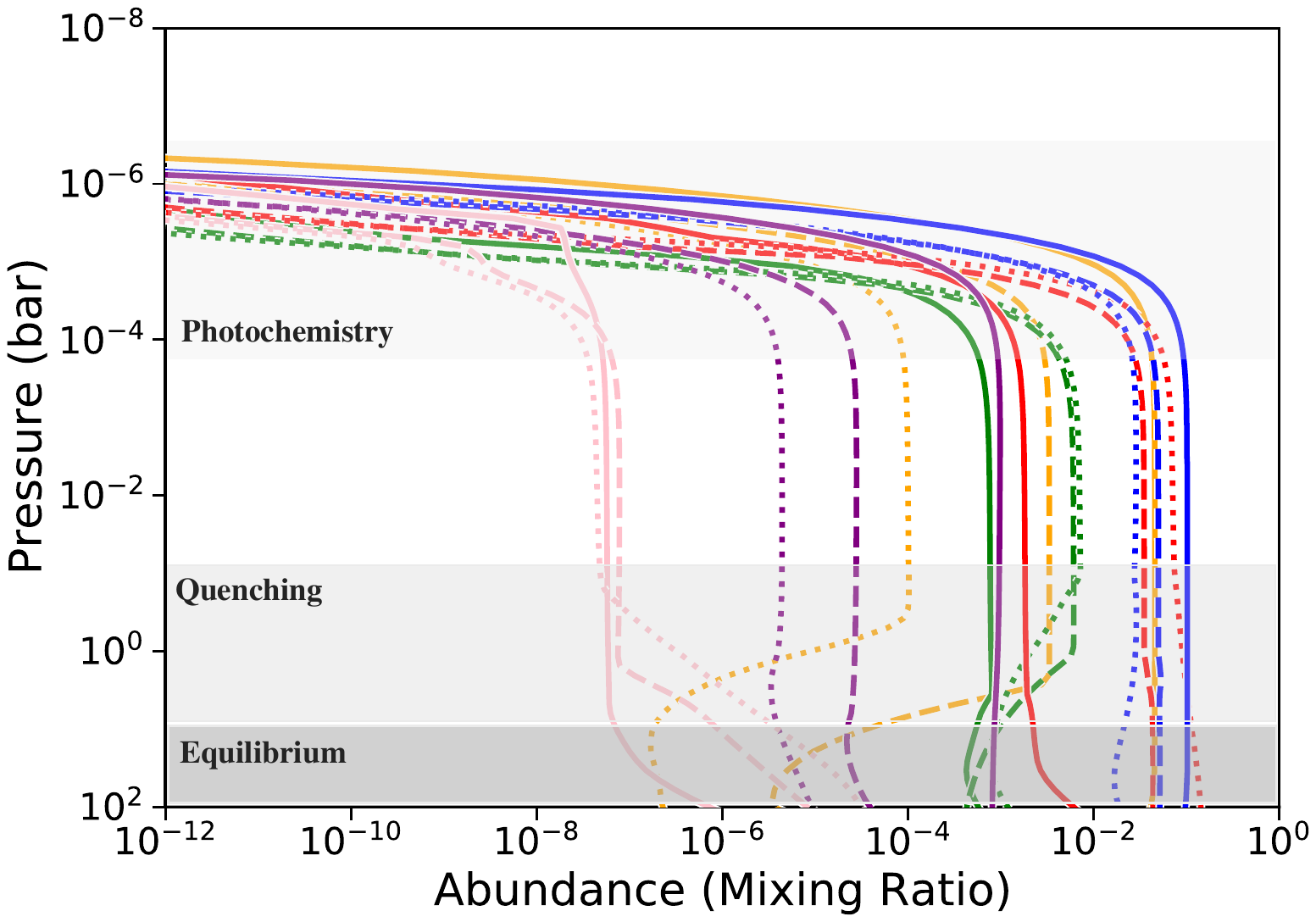}
    \subcaption{Constant $K_{\mathrm{zz}} = 10^{5}\ \mathrm{cm}^{2}\ \mathrm{s}^{-1}$}
    \label{fig:constkzz_1e5}
  \end{subfigure}\hfill
  \begin{subfigure}[t]{0.48\textwidth}
    \centering
    \includegraphics[width=\linewidth]{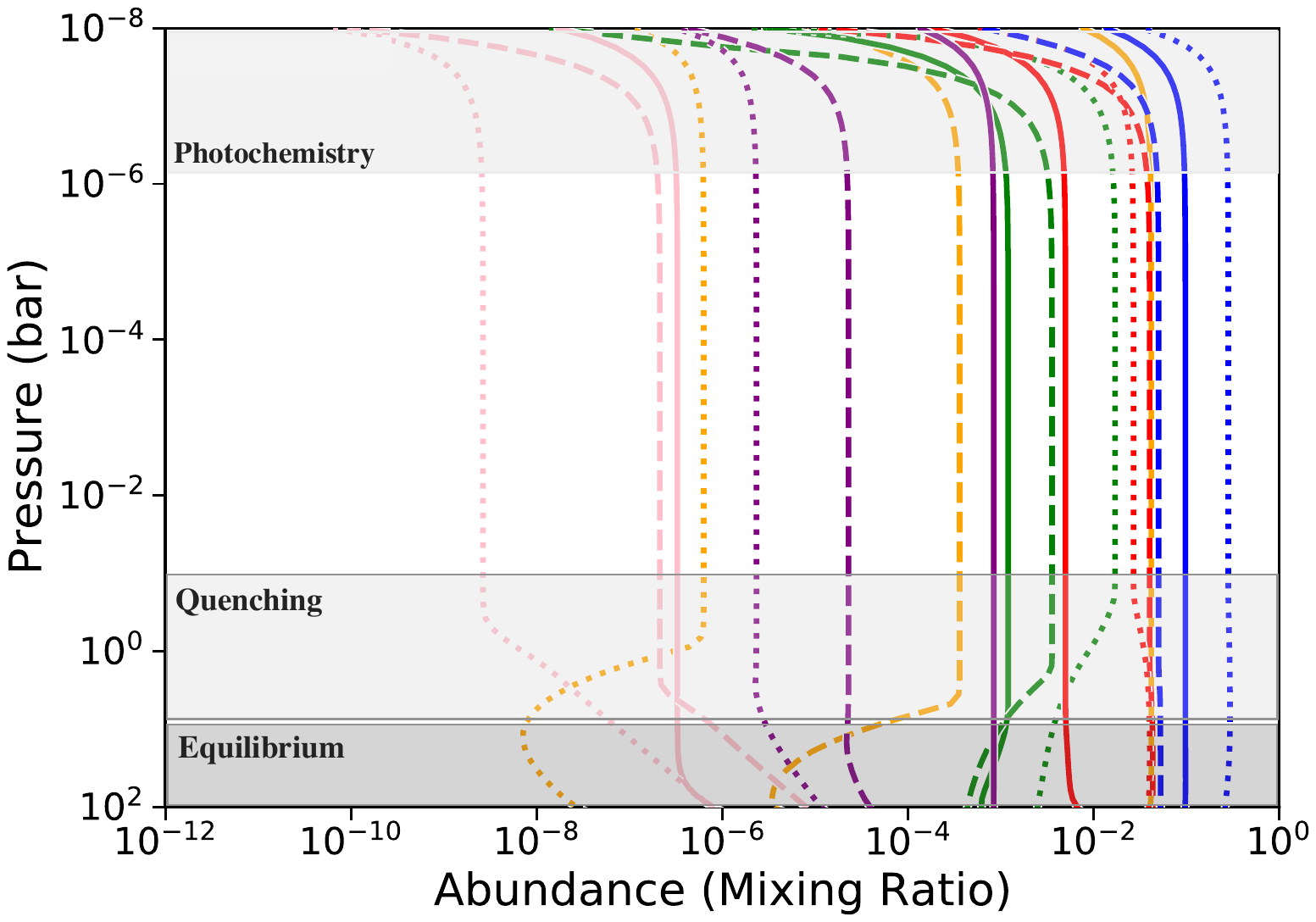}
    \subcaption{Constant $K_{\mathrm{zz}} = 10^{8}\ \mathrm{cm}^{2}\ \mathrm{s}^{-1}$}
    \label{fig:constkzz_1e8}
  \end{subfigure}

  \vspace{-1.5em}

  \begin{subfigure}[t]{0.48\textwidth}
    \centering
    \includegraphics[width=\linewidth]{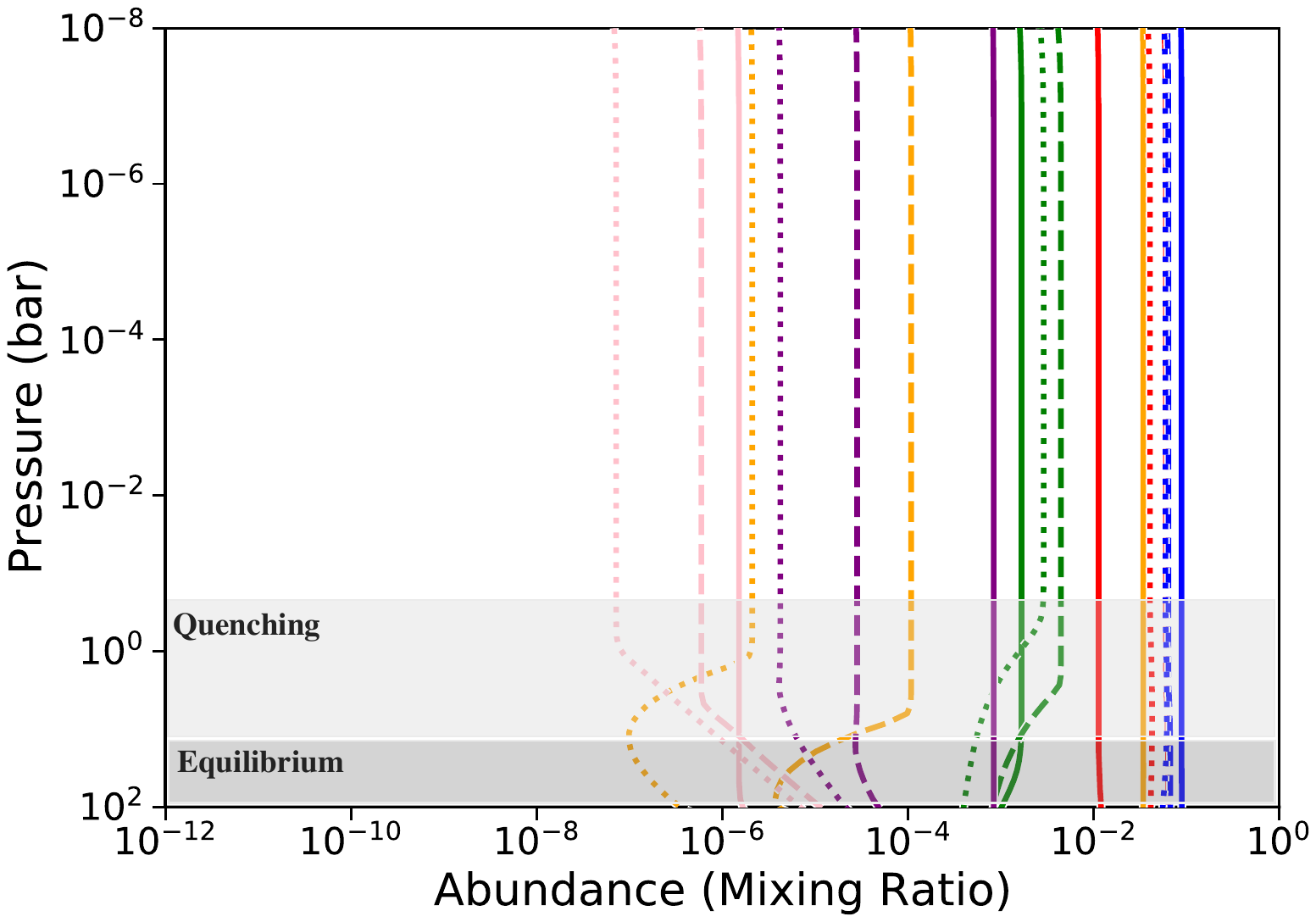}
    \subcaption{Constant $K_{\mathrm{zz}} = 10^{10}\ \mathrm{cm}^{2}\ \mathrm{s}^{-1}$}
    \label{fig:constkzz_1e10}
  \end{subfigure}\hfill
  \begin{subfigure}[t]{0.48\textwidth}
    \centering
    \includegraphics[width=\linewidth]{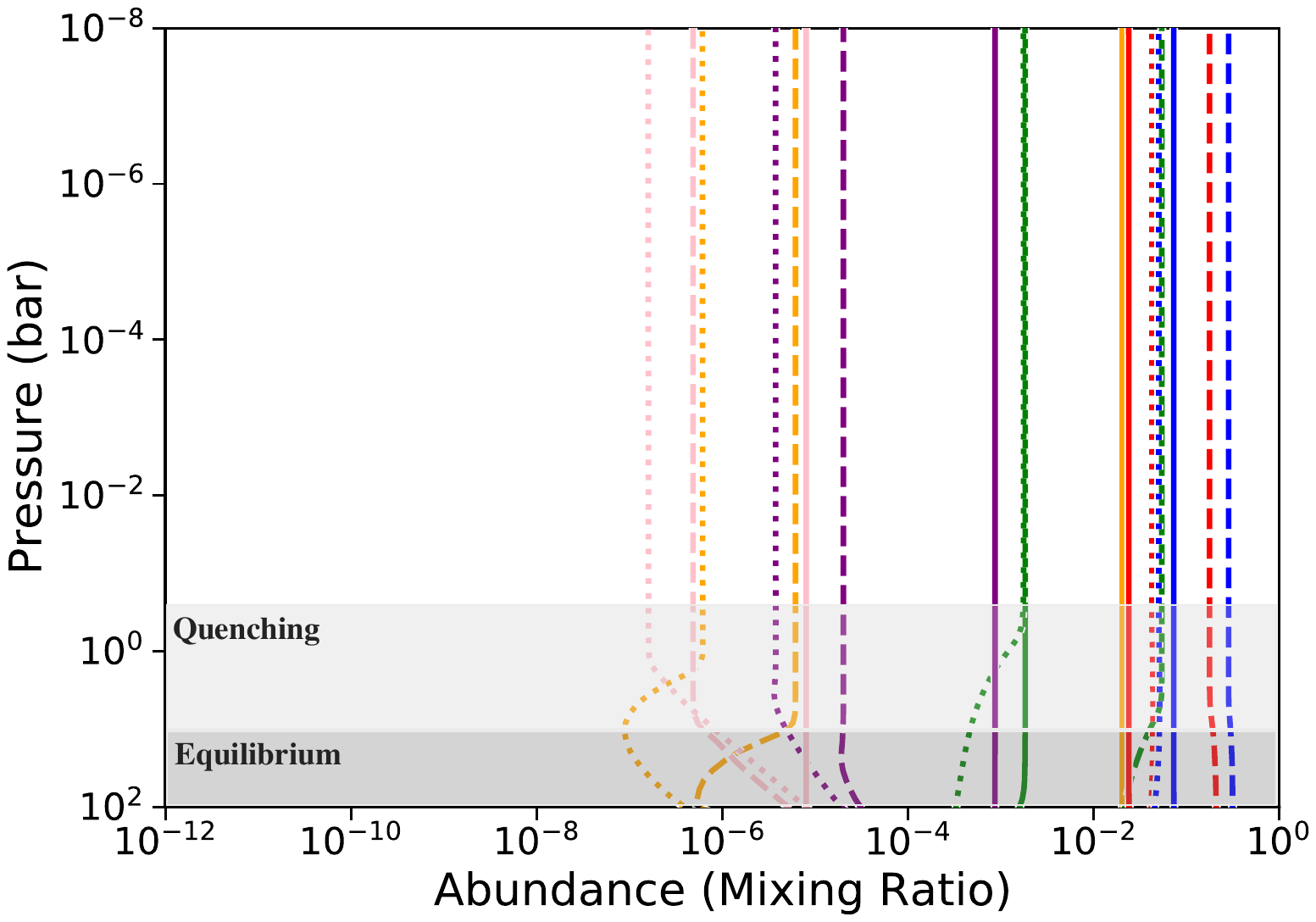}
    \subcaption{Constant $K_{\mathrm{zz}} = 10^{12}\ \mathrm{cm}^{2}\ \mathrm{s}^{-1}$}
    \label{fig:constkzz_1e12}
  \end{subfigure}

  \vspace{-1.5em}

  \begin{subfigure}[t]{0.48\textwidth}
    \centering
    \includegraphics[width=\linewidth]{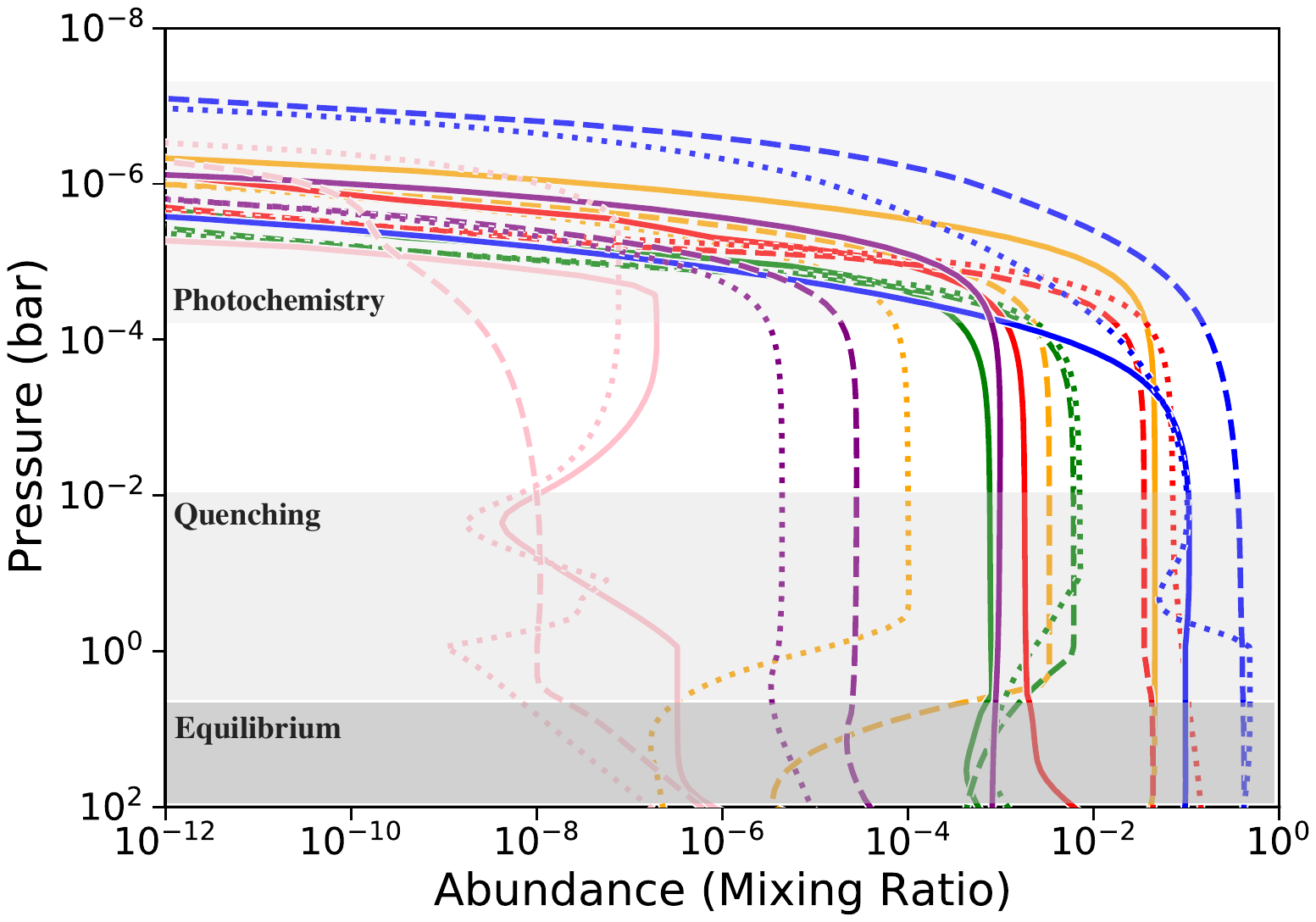}
    \subcaption{Variable $K_{\mathrm{zz}}$ profile}
    \label{fig:varkzz_profile}
  \end{subfigure}\hfill
  \begin{subfigure}[t]{0.48\textwidth}
    \centering
    \raisebox{1.5cm}{\includegraphics[width=0.7\linewidth]{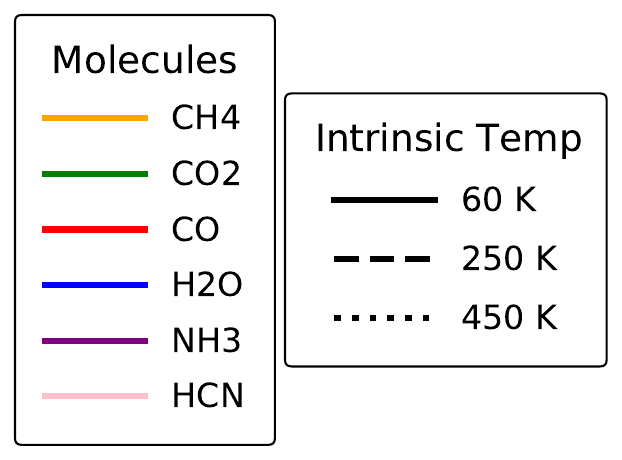}}
    \label{fig:abundance_sixth}
  \end{subfigure}

  \caption{Vertical abundance profiles of key atmospheric molecules as a function of pressure for mini-Neptune models with different eddy diffusion prescriptions. Panels (a)--(d) show constant $K_{\mathrm{zz}}$ cases: $10^{5}$, $10^{8}$, $10^{10}$, and $10^{12}\ \mathrm{cm}^{2}\ \mathrm{s}^{-1}$. Panel (e) shows the variable-$K_{\mathrm{zz}}$ case adopted from \citet{Hu2021} and \citet{Wogan2024}. Each panel shows mixing-ratio profiles for CH$_4$ (blue), CO$_2$ (green), CO (orange), H$_2$O (red), NH$_3$ (purple), and HCN (pink) at three intrinsic temperatures: $60$~K (solid lines), $250$~K (dashed lines), and $450$~K (dotted lines). The atmosphere is divided into three regions: the equilibrium zone (dark gray, $>100$~bar), the quenching zone (gray, $1$--$100$~bar), and the photochemistry zone (light gray, $<1$~bar).}
  \label{fig:abundance_profiles_all}
\end{figure}

Figure \ref{fig:abundance_profiles_all} reveals three distinct atmospheric regimes governing molecular distributions. In the deep equilibrium region ($\gtrsim 100$~bar), all molecules rapidly adjust to thermochemical equilibrium \citep{Visscher2006}. Quenching manifests as a characteristic ``kink'' in abundance profiles where chemical timescales exceed mixing timescales, followed by flat vertical profiles where quenched abundances are transported upward and remain constant until molecular diffusion dominates \citep{Visscher2010}.

The depth of quenching varies dramatically with mixing strength. For weak mixing ($10^{5}$ cm$^2$/s), quenching occurs between $10^{-1}$ to $10^{1}$~bar for most species. As mixing strengthens to $10^{8}$ cm$^2$/s, quench points deepen to $10^{-1}$--$1$~bar, and under extremely strong mixing ($10^{10}$ - $10^{12}$ cm$^2$/s), quenching shifts to $1$--$10$~bar. This systematic deepening enables high-temperature chemistry from deep layers to reach the observable photosphere ($10^{-3}$--$10^{-4}$~bar).

Different molecules exhibit distinct sensitivities to this quenching behavior, reflecting fundamentally different chemical processes. CH$_4$ and NH$_3$ show minimal quench-pressure variation across $K_{\mathrm{zz}}$ values, reflecting rapid equilibration timescales that keep pace with even vigorous vertical mixing \citep{Visscher2010}. For these thermochemically controlled species, the quench point preserves deep interior abundances by ``freezing'' them when transport becomes faster than chemistry.

HCN demonstrates extreme sensitivity, but through an opposite mechanism. Unlike CH$_4$ and NH$_3$, HCN is not formed in thermochemical equilibrium in the deep interior. Instead, it is photochemically produced in the upper atmosphere and then transported downward by mixing. The ``quench point'' for HCN represents the level where it is thermochemically destroyed and converted back to NH$_3$, CH$_4$, and other equilibrium species, rather than preserved \citep{Moses2011, Zahnle2009}. At high temperatures with weak mixing ($K_{\mathrm{zz}}$ = $10^{5}$ cm$^2$/s), this destruction occurs at shallow depths ($\sim 0.1$~bar), allowing HCN to accumulate in the observable atmosphere. Under strong mixing ($K_{\mathrm{zz}}$ = $10^{12}$ cm$^2$/s), vigorous transport carries HCN deeper into the hot interior where destruction is efficient, causing severe depletion despite continuous photochemical production above.

The variable-$K_{\mathrm{zz}}$ profile from \citet{Wogan2024} produces intermediate quenching behavior for equilibrium-dominated species (CH$_4$, NH$_3$, CO, H$_2$O). However, HCN experiences asymmetric effects: enhanced mixing at low pressures drives deeper quenching than expected, particularly affecting cool intrinsic models where the photosphere samples these strongly mixed upper layers. This altitude dependent mixing causes HCN to exhibit high $K_{\mathrm{zz}}$ like depletion in cool regimes despite weaker deep mixing \citep{Moses2011, Tsai2021}.

At pressures below $10^{-4}$~bar, abundance profiles deviate from quenched values as UV photolysis becomes dominant. This photochemical region is most prominent under weak mixing ($K_{\mathrm{zz}}$ = $10^{5}$ cm$^2$/s) and variable-$K_{\mathrm{zz}}$ conditions, where molecules like CH$_4$, H$_2$O, and NH$_3$ undergo photodissociation, creating radicals (CH$_3$, OH, NH$_2$) that drive secondary chemistry \citep{Moses2011, Hu2021}. Our treatment considers gas-phase photochemistry only and excludes photochemical haze formation and associated radiative effects. The photochemical layer thickness decreases with increasing $K_{\mathrm{zz}}$ as strong vertical transport creates a two-way exchange: upward transport continuously replenishes photolyzed parent molecules from below, while downward transport carries photochemical products into deeper layers where they are thermochemically converted back to their parent species. For $K_{\mathrm{zz}}$ $>$ $10^{10}$ cm$^2$/s, photochemical effects become negligible as mixing timescales far exceed destruction timescales, effectively creating a well-mixed atmosphere where photochemical signatures are diluted.

These vertical profiles demonstrate that observable atmospheric composition ($10^{-3}$--$10^{-4}$~bar) results from the combined influence of deep quenched abundances, condensation processes, and upper atmosphere photochemistry rather than local thermochemical equilibrium. The systematic $K_{\mathrm{zz}}$ and temperature dependence of quench pressures provides a diagnostic framework for constraining thermal structure ($T_{\mathrm{int}}$), mixing efficiency ($K_{\mathrm{zz}}$), and condensation efficiency through molecular abundance ratios, enabling us to break parameter degeneracies in atmospheric interpretation \citep{Tsai2021, Madhusudhan2023, Wogan2024}.

\subsection{Transmission Spectra} \label{sec: spectra}

\begin{figure}[!h]
    \centering
    \includegraphics[width=1.002\linewidth]{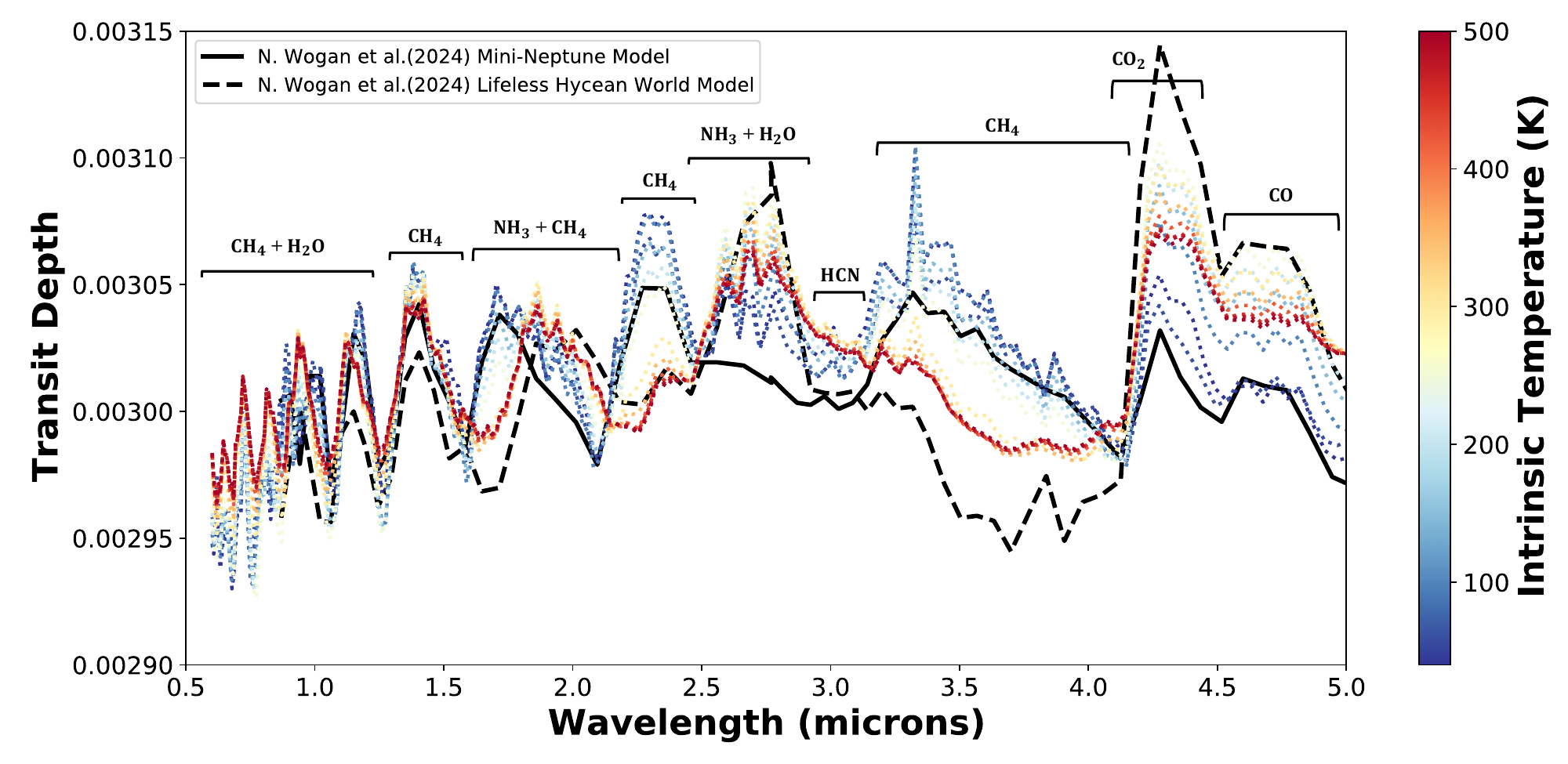}
    \caption{Spectrum grid with varying intrinsic temperature and \(K_{zz}\) profile. The \(K_{zz}\) profile follows the parameterization from \citet{Wogan2024} for mini-Neptune atmospheres. The grid showcases atmospheric temperature pressure profiles spanning intrinsic temperatures (\(T_\mathrm{int}\)) from 60 to 450~K with the \(K_{zz}\) profile applied across all models, while metallicity and C/O ratio are held constant. Note that the benchmark spectrum (black line) exhibits some artificially sharp features due to the spectral resolution and binning used in the original \citet{Wogan2024} analysis, which we maintain for direct comparison.}
    \label{fig:spectrum_tint}
\end{figure}

\begin{figure}[!h]
    \centering
    \includegraphics[width=1.002\linewidth]{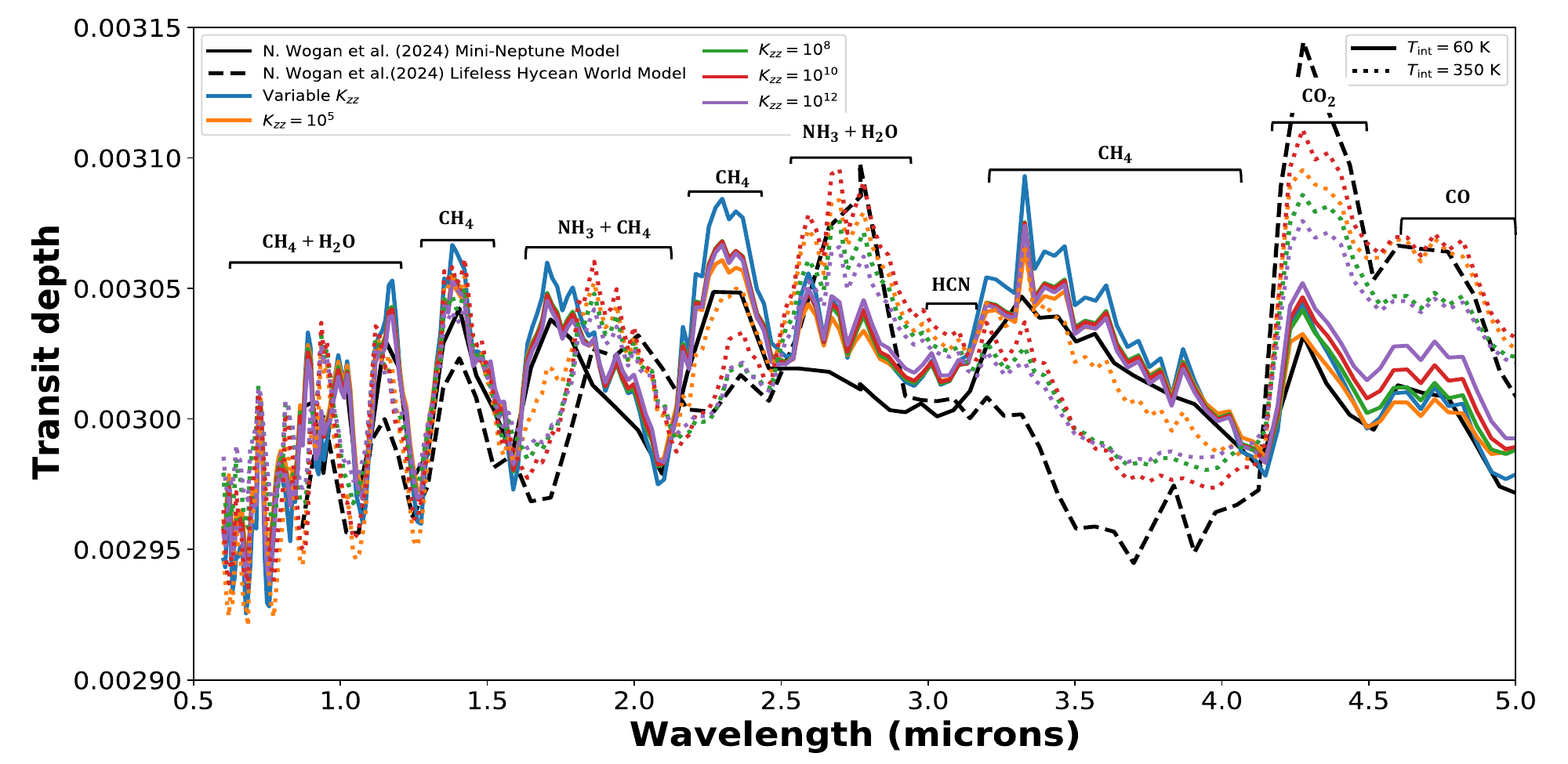}
    \caption{Transmission spectra for K2-18b analog atmospheres with fixed intrinsic temperature ($T_{\mathrm{int}} = 60$~K) and varying vertical eddy diffusion coefficients ($K_{\mathrm{zz}}$). The spectra span six orders of magnitude in mixing strength from $K_{zz}=10^{5}$ to $10^{12}~\mathrm{cm}^{2}\,\mathrm{s}^{-1}$ (colored lines), along with the variable $K_{\mathrm{zz}}$ profile from \citet{Wogan2024} (blue line). The \citet{Wogan2024} mini-Neptune benchmark model is shown in black for comparison.}
    \label{fig:spectrum_kzz}
\end{figure}

Figures \ref{fig:spectrum_tint} and \ref{fig:spectrum_kzz} present transmission spectra across our model grid from 0.5--5~$\mu$m, demonstrating how intrinsic temperature and vertical mixing control observable spectral signatures. The molecular abundance patterns discussed in Section \ref{sec:Tint+Kzz Effects} translate directly into characteristic spectral features that distinguish thermal regimes and mixing conditions.

Figure~\ref{fig:spectrum_tint} shows systematic spectral evolution with intrinsic temperature using the variable-$K_{\mathrm{zz}}$ profile. Cool $T_{\mathrm{int}}$ models ($T_{\mathrm{int}} \leq 150$~K) exhibit strong CH$_4$ features at $1.4$, $2.3$, and $3.3~\mu$m, accompanied by prominent NH$_3$+CH$_4$ absorption at $1.5$ and $2.0~\mu$m \citep{Fortney2005, Madhusudhan2023}. These signatures reflect the high CH$_4$ and NH$_3$ abundances produced by favorable thermochemical equilibrium at low temperatures  \citep{Visscher2006, Moses2011}. Intermediate models ($T_{\mathrm{int}} = 200$--$350$~K) show progressive weakening of CH$_4$ features with concurrent strengthening of CO$_2$ absorption at $4.3~\mu$m, capturing the transition from reducing to more oxidized atmospheric chemistry. Hot $T_{\mathrm{int}}$ models ($T_{\mathrm{int}} \geq 350$~K) display severely depleted CH$_4$ features across all diagnostic wavelengths, with dominant H$_2$O absorption and emerging CO features at $4.6~\mu$m reflecting the thermochemical dominance of oxidized carbon species \citep{Moses2011, Tsai2021}.

Crucially, our intermediate models ($T_{\mathrm{int}} = 300$--$350$~K) show systematic convergence toward the \citet{Wogan2024} sub-Neptune benchmark, reaching a difference of $34$~ppm across $0.9$--$5.1~\mu$m. The agreement varies by molecular region: excellent convergence occurs for the primary CH$_4$ diagnostic band at $3.3~\mu$m ($10$~ppm), while larger systematic offsets appear in $2.5$--$3.0~\mu$m; ($53$~ppm) and CO$_2$/CO bands ($4.3$--$4.6~\mu$m; $44$--$66$~ppm). Despite these localized discrepancies, the overall spectral morphology and dominant molecular absorption signatures remain consistent, indicating that this thermal regime reproduces the characteristic sub-Neptune spectrum while capturing subtle differences in atmospheric chemistry parameterization. This convergence occurs precisely where our abundance analysis indicates favorable CO$_2$/CH$_4$ ratios and balanced thermochemistry and transport interactions.

Figure~\ref{fig:spectrum_kzz} reveals the contrasting role of vertical mixing at fixed cool temperature ($T_{\mathrm{int}} = 60$~K). Despite spanning six orders of magnitude in $K_{\mathrm{zz}}$ ($10^{5}$ to $10^{12}~\mathrm{cm}^{2}\,\mathrm{s}^{-1}$), the transmission spectra show similarity in their primary molecular features. The CH$_4$ absorption bands at $1.4$, $2.3$, and $3.3~\mu$m remain virtually unchanged across all mixing strengths, as do the NH$_3$ features at $1.5$ and $2.0~\mu$m \citep{Fortney2005, Madhusudhan2023}. This spectral insensitivity confirms that at cool temperatures, thermochemical equilibrium dominates over transport effects, with molecular abundances controlled by local temperature rather than vertical mixing efficiency \citep{Visscher2006, Visscher2010, Moses2011}.

However, subtle differences do emerge in specific wavelength regions. The CO$_2$ feature at $4.3~\mu$m shows modest variation with $K_{\mathrm{zz}}$, reflecting the moderate mixing sensitivity identified in our abundance analysis. Additionally, the variable-$K_{\mathrm{zz}}$ profile from \citet{Wogan2024} produces spectra that fall within the range of constant-mixing models, validating our systematic exploration approach. The convergence of all cool-temperature spectra toward similar morphologies, regardless of mixing strength, highlights that thermal state, not atmospheric dynamics, provides the primary spectral diagnostic for mini-Neptune characterization.

These systematic spectral trends and benchmark convergence provide observational constraints on atmospheric thermal state and mixing efficiency, establishing a framework for distinguishing between competing atmospheric scenarios through transmission spectroscopy, with detailed diagnostic implications discussed in Section~\ref{sec: Discussion}.

\section{Discussion} \label{sec: Discussion}

Our systematic exploration of intrinsic temperature and vertical mixing parameter space reveals fundamental insights that extend and complement previous sub-Neptune atmospheric studies. While earlier investigations focused primarily on surface depth effects \citep[e.g.,][]{Yu2021}, atmospheric metallicity variations \citep[e.g.,][]{Jaziri2025}, and stellar irradiation impacts \citep[e.g.,][]{Hu2025}, our work demonstrates that interior thermal state and vertical transport parameters typically held fixed, exert comparable control over observable atmospheric composition. The convergence of our intermediate-temperature models ($T_{\mathrm{int}} = 300$--$350$~K) with the \citet{Wogan2024} K2-18b benchmark establishes that moderately heated sub-Neptunes can reproduce observed spectral signatures through deep thermochemical processes alone, creating substantial degeneracy with shallow-surface Hycean world scenarios. This finding challenges the conventional approach of inferring planetary composition from atmospheric spectra without systematic exploration of thermal and mixing parameter space. The distinct sensitivity regimes we identify, from temperature-dominated species (CH$_4$, NH$_3$) to extreme disequilibrium tracers (HCN), provide new diagnostic frameworks for distinguishing between competing atmospheric scenarios. These results have direct implications for interpreting current JWST observations, designing future observational strategies, and reassessing the census of potentially habitable worlds among the sub-Neptune population.

\subsection{Justification for Parameter Space}\label{subsec: hotinterior}

Our systematic exploration reveals that intrinsic temperatures ranging from
$T_{\mathrm{int}} = 60$--$450~\mathrm{K}$ produce significant changes in the retrieved molecular abundances and observable transmission spectrum (Figure \ref{fig:abundances-six}). We note that the extreme range in intrinsic temperatures is plausible for sub-Neptunes. Gravitational contraction represents a primary heat source, with cooling timescales extending to $1~\mathrm{Gyr}$ for planets in the $2$--$8~M_{\oplus}$ mass range \citep{Fortney2020}. Radiogenic heating from long-lived isotopes (e.g., $^{26}\mathrm{Al}$, $^{40}\mathrm{K}$, U, Th) can also provide sustained interior heat over geological timescales \citep{Lammer2018}, while the retention of substantial H/He envelopes impedes radiative heat loss and can maintain elevated intrinsic temperatures over extended periods \citep{Lopez2014}. Tidal heating from eccentric or obliquity tides can further dump heat into the interior of the planet throughout its lifetime, enabling even older planets to maintain heated interiors \citep{Millholland2019, Millholland2020}. Recent observations of the more massive sub-Saturn, WASP-107b, determine a $T_{\mathrm{int}}$ of $460 \pm 40~\mathrm{K}$, an artifact of ongoing tidal heating \citep{Sing2024}.

On the other hand, older planets not experiencing an ongoing source of heat are expected to cool over time. Atmospheric models \citep{Fortney2011, Guillot2010, Thorngren2019} often assume $T_{\mathrm{int}}$ between $40$--$100~\mathrm{K}$ for planets older than several billion years. However, even in our own Solar System, it is unclear what scenarios control the $T_{\mathrm{int}}$ of our gas giants. Neptune outputs nearly $3\times$ more internal heat than Uranus, despite their similarities in size and composition \citep{Guillot2015}. Therefore, without knowing the evolutionary history of sub-Neptunes, exploring the full range of plausible intrinsic temperatures is required for deriving atmospheric composition and structure.

While less of a driver of disequilibrium chemistry, $K_{\mathrm{zz}}$ impacts the quench pressure level and overall mixing of the observable pressures in an atmosphere. Observational constraints, however, are limited. Measured $K_{\mathrm{zz}}$ in our Solar System gas giants vary from $10^{6}$ to $10^{9}~\mathrm{cm^{2}\,s^{-1}}$ \citep{Moses2011, Zhang2018}, while theoretical models suggest a range of $K_{\mathrm{zz}}$ from $10^{5}$ to $10^{12}~\mathrm{cm^{2}\,s^{-1}}$ \citep{Parmentier2013, Zhang2018}. Given the limited observational constraints on the vertical mixing of sub-Neptunes, we explored a wide range of constant $K_{\mathrm{zz}}$ profiles alongside the variable $K_{\mathrm{zz}}$ profile adopted from \citet{Wogan2024}. Comparing our transmission spectra across these different mixing assumptions reveals remarkably similar observable signatures. The variable $K_{\mathrm{zz}}$ profile produces spectral features that closely match those from constant $K_{\mathrm{zz}}$ cases in the $10^{8}$--$10^{10}~\mathrm{cm^{2}\,s^{-1}}$ range, with differences of less than $5$~ppm in key diagnostic wavelengths ($1.4$, $2.3$, and $3.3~\mu\mathrm{m}$ for CH$_4$; $4.3~\mu\mathrm{m}$ for CO$_2$). Even for molecules most sensitive to mixing strength, such as HCN, the variable profile behavior falls within the envelope of constant $K_{\mathrm{zz}}$ predictions rather than representing a fundamentally distinct regime.

This spectral similarity demonstrates that for atmospheric characterization purposes, assuming a constant $K_{\mathrm{zz}}$ profile provides a robust approximation to more complex, altitude-dependent mixing scenarios. The choice between constant and variable $K_{\mathrm{zz}}$ parameterizations introduces uncertainties smaller than current observational precision, validating the use of constant $K_{\mathrm{zz}}$ grids for interpreting sub-Neptune transmission spectra. This finding is particularly important for atmospheric retrievals, where computational efficiency often requires simplified mixing prescriptions.

\subsection{In Context with Previous sub-Neptune Work}\label{subsec: prevwork}

Our approach draws from multiple atmospheric modeling frameworks to understand compositional degeneracies in sub-Neptune atmospheres. In particular, our work builds on the framework developed by \citet{Yu2021}, who showed that surface depth controls how molecules behave in sub-Neptune atmospheres. In their model, deep atmospheres without surfaces allow CH$_4$ and NH$_3$ to be recycled: these molecules are destroyed by starlight in the upper atmosphere, but vertical mixing carries them down to hot deep layers where they reform, and then transports them back upward. This recycling process prevents molecular abundances from being completely eroded by stellar radiation.

However, shallow surfaces at around 10 bar pressure cut off this recycling. In Hycean world scenarios, a water ocean surface stops the atmosphere early, so destroyed CH$_4$ and NH$_3$ molecules cannot reach the deep hot layers to be reformed \citep{Yu2021}. Instead, these molecules are continuously destroyed without replacement, leading to depleted abundances and more oxidized chemistry with CO and CO$_2$.

\begin{figure}[!h]
  \centering
  \begin{subfigure}{0.24\textwidth}
    \includegraphics[width=1.01\linewidth]{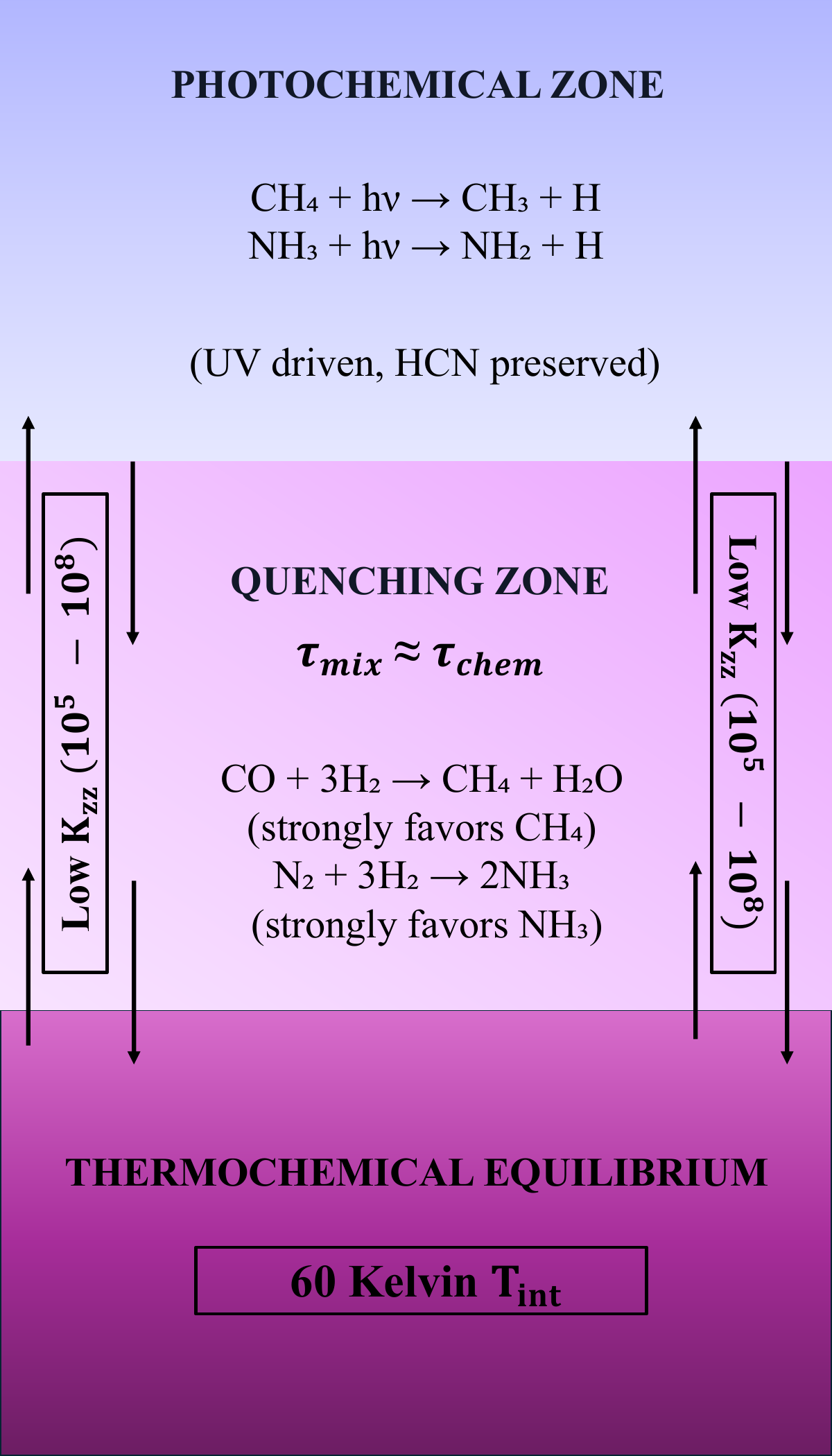}
    \vspace{3pt}
    \caption{Low T$_{\rm int}$, Low K$_{zz}$}
    \label{fig:lowtint_lowkzz}
  \end{subfigure}\hfill
  \begin{subfigure}{0.24\textwidth}
    \includegraphics[width=1.01\linewidth]{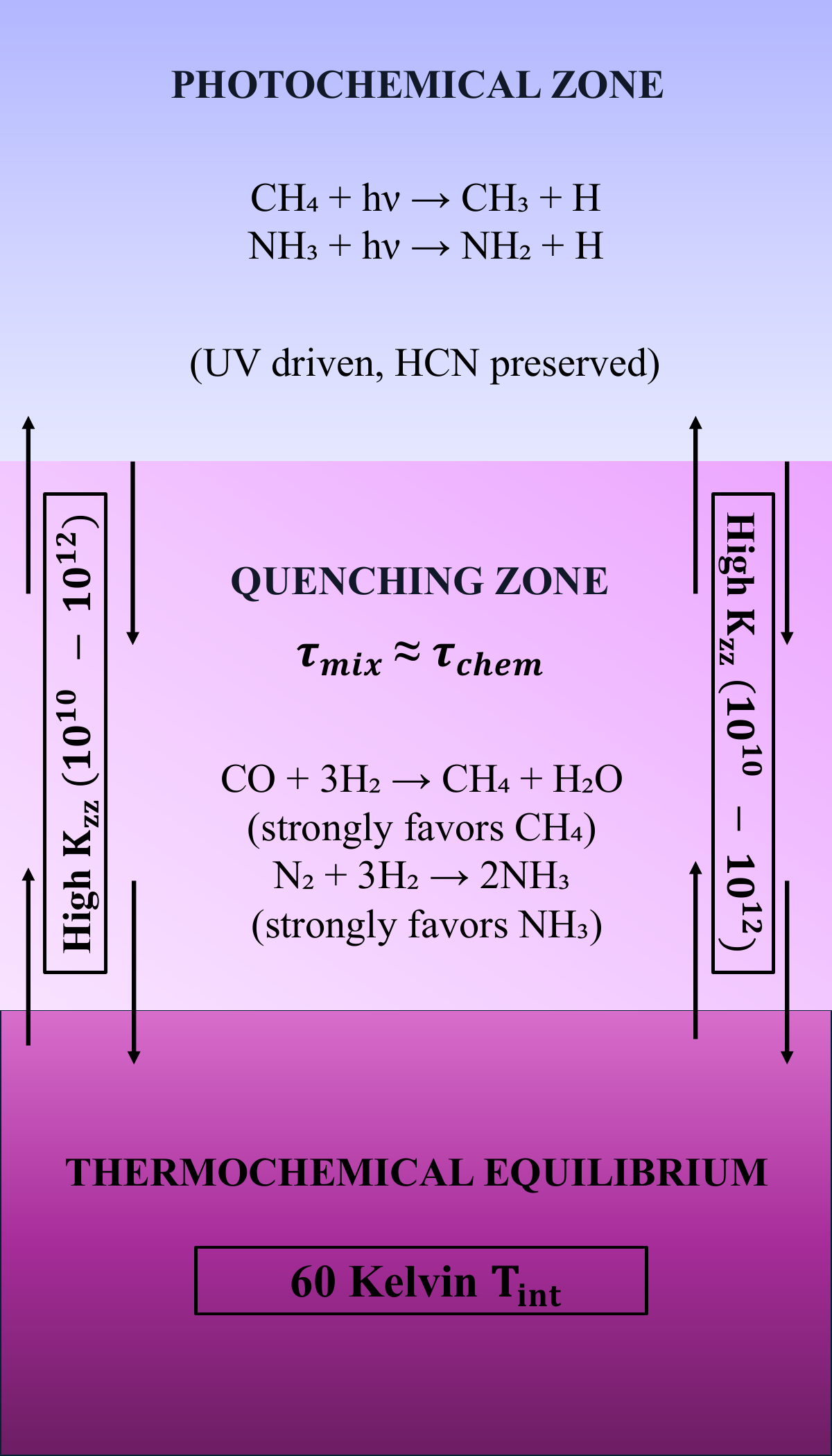}
    \vspace{3pt}
    \caption{Low T$_{\rm int}$, High K$_{zz}$}
    \label{fig:lowtint_highkzz}
  \end{subfigure}\hfill
  \begin{subfigure}{0.24\textwidth}
    \includegraphics[width=1.01\linewidth]{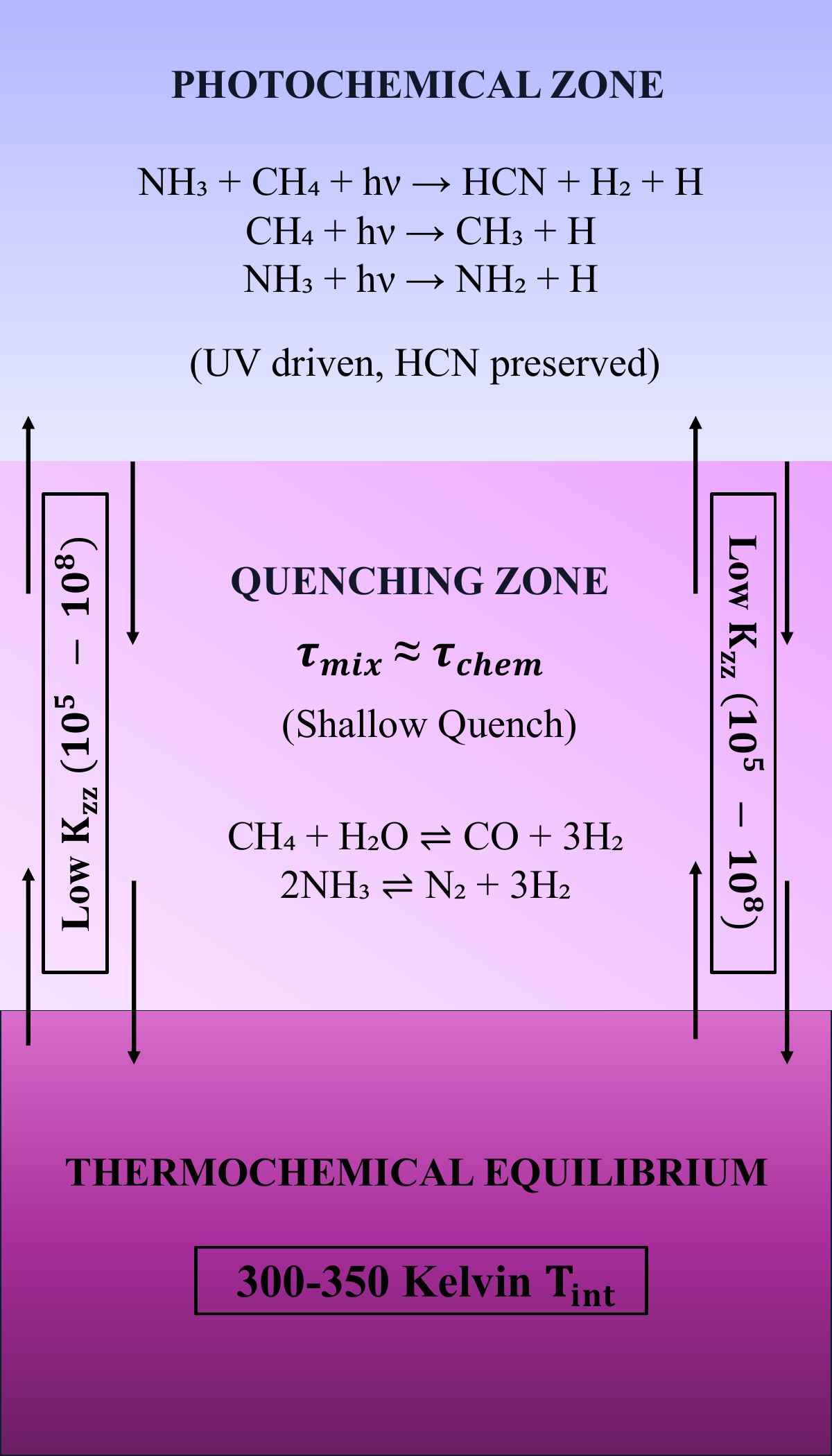}
    \vspace{3pt}
    \caption{Int. T$_{\rm int}$, Low K$_{zz}$}
    \label{fig:hightint_lowkzz}
  \end{subfigure}\hfill
  \begin{subfigure}{0.24\textwidth}
    \includegraphics[width=1.01\linewidth]{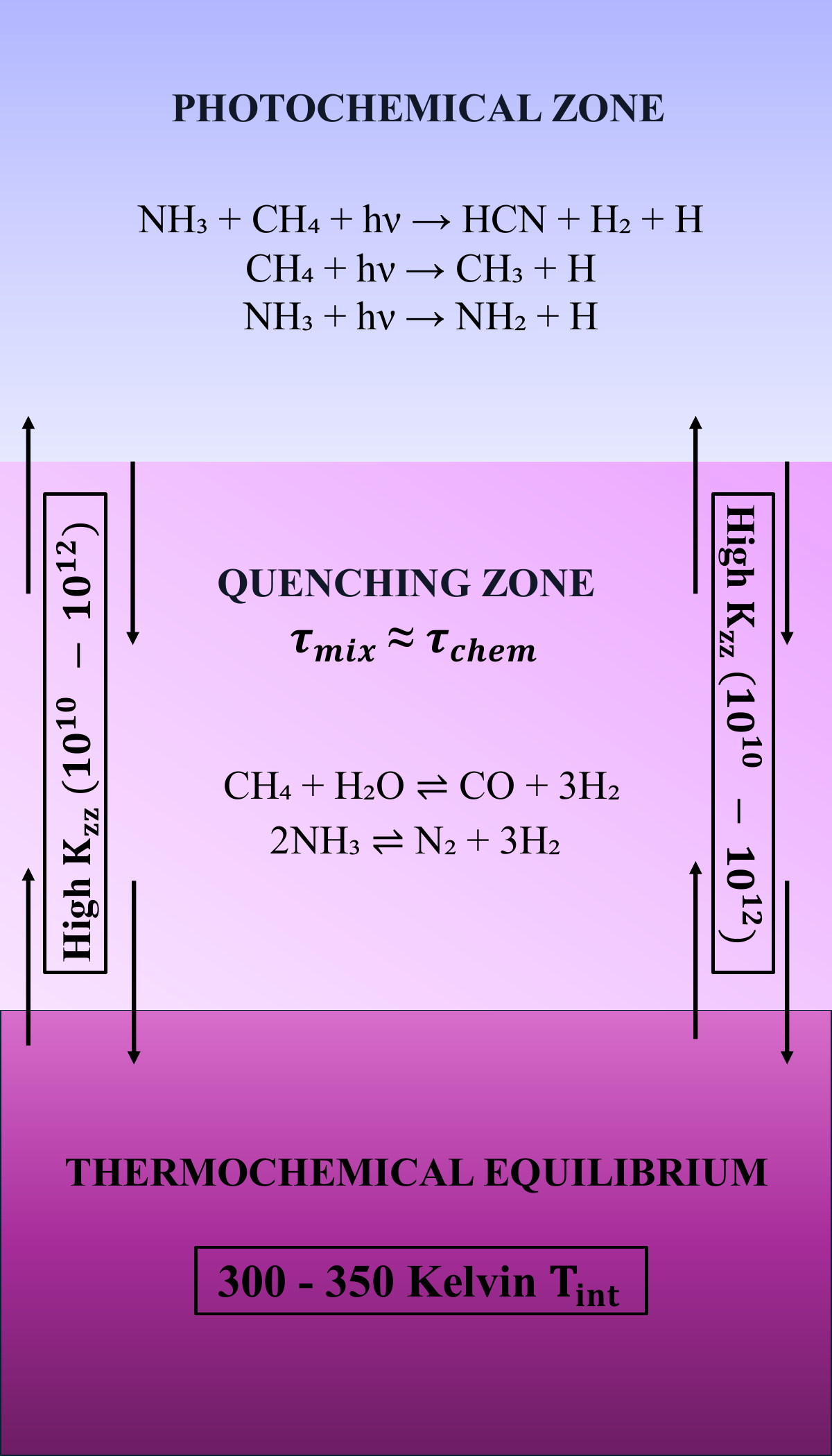}
    \vspace{3pt}
    \caption{Int. T$_{\rm int}$, High K$_{zz}$}
    \label{fig:hightint_highkzz}
  \end{subfigure}
  \caption{Schematic illustration of atmospheric chemistry regimes for different intrinsic temperature and vertical mixing combinations in K2-18b analogs. The atmosphere is divided into three regions: the equilibrium zone (purple, deep atmosphere), the quenching zone (pink, where $\tau_{\rm mix} \approx \tau_{\rm chem}$), and the photochemistry zone (blue, upper atmosphere). (a) Cool interior with weak vertical mixing (b) Cool interior with strong vertical mixing (c) Intermediate temperature with weak vertical mixing (d) Intermediate temperature with strong vertical mixing}
  \label{fig:discussion plots}
\end{figure}

Our quantitative analysis reveals that conditions for reproducing K2-18b's molecular signatures occur when chemical ($\tau_{\mathrm{chem}}$) and mixing timescales,
\begin{equation}
\tau_{\mathrm{mix}}=\frac{L^{2}}{K_{zz}},
\end{equation}
converge under intermediate intrinsic temperatures ($T_{\mathrm{int}}=250$--$350$~K; $K_{zz}=10^{8}$--$10^{10}~\mathrm{cm}^{2}\,\mathrm{s}^{-1}$). For CH$_4$, chemical timescales accelerate by a factor of ten from cool to hot conditions, while mixing timescales span four orders of magnitude across our $K_{\mathrm{zz}}$ range. This convergence deepens quench pressures from $10^{8}$~bar to $10^{7}$~bar, enabling transport of interior chemistry to observable altitudes, analogous to quenching behavior in Jupiter's atmosphere \citep{Moses2011}.

Figure~\ref{fig:discussion plots} illustrates how these processes combine in K2-18b analogs. Cool interior models consistently produce CH$_4$-rich atmospheres regardless of mixing strength, failing to match observations. However, intermediate-temperature models with strong mixing achieve the moderate CH$_4$ depletion and CO$_2$ enhancement observed by JWST \citep{Madhusudhan2023}. Remarkably, this produces molecular patterns similar to Hycean predictions but through fundamentally different mechanisms: deep-interior processes in thick atmospheres rather than surface--ocean chemistry in thin ones. Our transmission spectra demonstrate this degeneracy directly. Models with intermediate intrinsic temperatures ($T_{\mathrm{int}} = 300$--$350$~K) closely reproduce the \citet{Wogan2024} K2-18b benchmark spectrum with differences of only 14.5~ppm across $0.5$--$5~\mu\mathrm{m}$, despite representing entirely different planetary scenarios. The spectral convergence occurs because both pathways can produce similar CH$_4$ depletion and CO$_2$ enhancement: Hycean worlds achieve this through photochemical destruction at shallow surfaces with no recycling, while our heated sub-Neptunes reach the same observable outcome through thermochemical quenching in deep, hot interiors transported upward by vertical mixing. This atmospheric complexity where systematic variations in just two parameters ($T_{\mathrm{int}}$ and $K_{\mathrm{zz}}$) can reproduce the spectral signatures previously attributed to fundamentally different planet types reveals that transmission spectroscopy alone may be insufficient to distinguish between competing atmospheric scenarios without additional diagnostics.

However, our parameter study reveals diagnostic opportunities through mixing sensitive molecular tracers. Species that form in deep atmospheric layers but face destruction in upper regions provide the clearest tests while some atmospheric scenarios might preserve these tracers, vigorous vertical transport can carry them to regions where chemical loss dominates (Example: HCN) \citep{Moses2011, Tsai2021}. The specific conditions we identify ($T_{\mathrm{int}}=300$--$350$~K with strong mixing) provide concrete observational targets for constraining planetary interiors and atmospheric dynamics \citep{Zhang2018}. Our results demonstrate that interpreting sub-Neptune atmospheres requires considering both surface conditions and interior thermal evolution. Future work should systematically explore how these processes combine across the full range of sub-Neptune properties to better understand this abundant planetary class \citep{Fulton2017, Rogers2015}.

\subsection{Constraints of this Work}\label{subsec: limitations}

This work presents a comprehensive analysis of the intrinsic temperature and vertical mixing coefficient parameter space, we acknowledge that assumptions made do not capture the full range of possibilities for characterizing the complexities of sub-Neptune atmospheres. While observationally supported for K2-18b and other sub-Neptunes, we hold the atmospheric composition constant to a 100$\times$ solar metallicity and a solar C/O ratio, which constrains the chemical diversity we can explore. Studies such as \citet{Jaziri2025} have demonstrated that atmospheric metallicity can significantly affect the observable spectra and molecular abundances in sub-Neptune atmospheres, we have not systematically varied this parameter. Additionally, the C/O ratio plays a crucial role in determining atmospheric composition by affecting the chemical equilibrium partitioning between carbon and oxygen bearing species. C/O ratios above or below solar values can substantially alter the relative abundances of key molecules such as $\text{CH}_4$, $\text{H}_2$O, $\text{CO}$, and $\text{CO}_2$, which are primary observational molecules \citep{Madhusudhan2012}. The interplay between metallicity, C/O ratio, and our explored parameters intrinsic temperature and vertical mixing likely creates a more complex parameter space than we have yet to investigate. Future work incorporating a grid of metallicity and C/O values would provide a more complete picture of the degeneracies and unique atmospheric signatures accessible through observations.

Our chemical network focuses on carbon, hydrogen, nitrogen, and oxygen (NCHO) species, excluding sulfur chemistry from our systematic parameter-grid exploration. While we verified that sulfur inclusion produces minimal differences in our benchmark models (1.8~ppm difference), sulfur-bearing species such as H$_2$S, SO$_2$, and other sulfur compounds can play important roles in exoplanet atmospheric chemistry and may exhibit distinct sensitivities to intrinsic temperature and vertical mixing that we have not systematically characterized. The recent detection and modeling of sulfur species in exoplanet atmospheres, such as the analysis of GJ 436b by \citet{Tsai2023}, demonstrates the potential observational significance of these molecules for atmospheric characterization. Future work should extend our systematic parameter space exploration to include sulfur chemistry networks, particularly for planets where sulfur species may be observable with JWST and other next generation instruments.

We do not explicitly include aerosol and haze production, which are known to significantly affect transmission spectra by reducing molecular absorption features and steepening spectral trends across wavelength \citep{Fortney2005, Sing2016}. In extreme cases, thick photochemical hazes can completely mute spectral features, as observed in planets like GJ~1214b \citep{Kempton2014}. While our intrinsic temperature variations account for some temperature-dependent processes, specifically water condensation and cold trapping, we do not model the full microphysical properties, vertical distributions, or compositional variations of condensate clouds for other molecular species, nor do we include photochemical hazes. However, our systematic exploration of intrinsic temperature effects provides insights into potential haze formation trends. Since photochemical hazes in sub-Neptune atmospheres are primarily produced from hydrocarbon precursors like CH$_4$ \citep{Hu2025}, our finding that higher intrinsic temperatures systematically deplete CH$_4$ abundances suggests an intriguing hypothesis: intrinsically hotter sub-Neptunes should exhibit clearer atmospheres due to reduced availability of haze precursors. This prediction could be tested observationally, as sub-Neptunes with evidence for heated interiors should show stronger molecular absorption features compared to their
cooler counterparts. While a full exploration of the complex coupling between intrinsic temperature, atmospheric chemistry, and haze microphysics represents an important avenue for future work, the fundamental chemical trends we identify provide a framework for understanding how interior thermal states may influence observable atmospheric properties beyond gas-phase chemistry alone. For K2-18b specifically, the lack of substantial haze signatures in current observations \citep{Madhusudhan2023} may be consistent with our inference of moderately heated interior conditions that deplete haze-forming precursors.

Photochemically active species such as NH$_3$ and CH$_4$ are particularly sensitive to the stellar UV environment, as photodissociation rates depend strongly on the stellar spectrum in the ultraviolet \citep{Hu2012, Moses2011}. Different host stars exhibit distinct UV to visible flux ratios, which can lead to different photochemical pathways and molecular abundances \citep{Segura2005, France2013, Kawashima2019}. \citet{Cooke2024} conducted a detailed exploration of K2-18's stellar spectrum and its photochemical implications for atmospheric modeling in their appendix, demonstrating the importance of accurate stellar characterization for sub-Neptune atmospheric chemistry. By restricting our analysis to a single stellar spectrum, we did not assess how variations in stellar characteristics affect the observability and abundance patterns of key atmospheric molecules in sub-Neptune atmospheres. Additionally, we do not account for stellar contamination effects in our transmission spectrum calculations. K2-18 exhibits higher activity than typical M dwarfs of its age due to observable star spots and activity cycles \citep{Barclay2021}, and such stellar contamination can introduce spectral features that mimic or mask atmospheric signatures \citep{Rackham2018, Rackham2019}. The enhanced activity of K2-18 emphasizes the importance of obtaining detailed UV spectra for individual host stars rather than relying on proxy spectra, as stellar activity affects both the photochemical environment and observed transmission spectra.

\section{Conclusion}\label{sec:conclusion}

We present the first comprehensive exploration of intrinsic temperature ($T_{\mathrm{int}}$) vertical mixing ($K_{\mathrm{zz}}$) parameter space for temperate sub-Neptune atmospheres (K2-18b analogs), spanning $T_{\mathrm{int}}=60$--450~K and $K_{zz}=10^{5}$--$10^{12}~\mathrm{cm}^{2}\,\mathrm{s}^{-1}$ using coupled \texttt{PICASO} and \texttt{VULCAN} models. Our systematic analysis reveals that these parameters, often treated as fixed in prior work, exert fundamental control over atmospheric composition and observable spectra.

\begin{enumerate}
    \item Molecular abundance patterns exhibit distinct sensitivity regimes. CH$_4$ and NH$_3$ show strong temperature dependence with comparatively weaker sensitivity to mixing strength, consistent with thermochemical control. CO and H$_2$O display moderate coupling between thermal and dynamical effects, while HCN demonstrates extreme disequilibrium behavior, varying by up to 6 orders of magnitude across the explored $K_{\mathrm{zz}}$ range. CO$_2$ emerges as a diagnostic of intermediate thermal states, reaching peak abundances at $T_{\mathrm{int}}=200$--350~K under moderate mixing conditions.

    \item Transmission spectra across 0.5--5~$\mu$m reveal systematic evolution with intrinsic temperature. Cool models ($T_{\mathrm{int}}<150$~K) produce excessive CH$_4$ absorption inconsistent with observations, while hot models ($T_{\mathrm{int}}>400$~K) show strongly depleted CH$_4$ and spectra dominated by H$_2$O opacity. Critically, intermediate models ($T_{\mathrm{int}}=250$--350~K) reproduce the K2-18b benchmark spectrum, demonstrating that moderate interior heating can explain the observed molecular patterns without invoking Hycean scenarios.

    \item Quenching analysis demonstrates that observable atmospheric composition reflects deep interior chemistry transported upward through pressure-dependent mixing. Chemical timescales decrease from $10^{6}$~s at low temperatures to $10^{4}$~s at high temperatures, converging with mixing timescales at intermediate $T_{\mathrm{int}}$ to yield the molecular abundance ratios observed in temperate sub-Neptunes.

    \item A key challenge for compositional interpretation is the degeneracy between gas-rich sub-Neptunes with hot interiors and Hycean worlds with shallow surfaces. Both scenarios can produce similar CH$_4$ depletion and CO$_2$ enhancement, but through fundamentally different mechanisms: deep quenching versus photochemical destruction without recycling. Our models provide a diagnostic framework using HCN detection, H$_2$O measurements, and NH$_3$ upper limits to distinguish these competing scenarios.

    \item For K2-18b specifically, spectral matching constrains the intrinsic temperature to $T_{\mathrm{int}} = 250$--350~K, indicating either a moderately heated sub-Neptune interior or alternative thermal structures consistent with different planetary scenarios. The convergence of our intermediate-temperature models with observed spectral features demonstrates that systematic exploration of thermal and mixing parameter space is essential for robust atmospheric interpretation, as compositional degeneracies can lead to misclassification of planetary scenarios without comprehensive parameter studies.

\end{enumerate}

These results fundamentally change approaches to sub-Neptune characterization by demonstrating that thermal and mixing parameters cannot be treated as fixed quantities. The substantial degeneracies we identify between different atmospheric models highlight that single parameter assumptions can lead to incorrect conclusions about planetary composition and habitability. For current JWST observations, this motivates prioritizing multi-molecule diagnostics over single species detections.

At the population level, our results reveal the complexity of inferring surface conditions from atmospheric observations alone. The systematic parameter dependencies we identify demonstrate that atmospheric composition reflects a complex interplay of interior thermal evolution, mixing processes, and surface conditions. This complexity suggests that robust characterization of sub-Neptune chemistry requires comprehensive parameter space exploration rather than relying on simplified atmospheric models with fixed assumptions.

Future work should extend this analysis to include aerosol and haze formation, systematic variations in atmospheric metallicity and C/O ratios, and quantify the impact of stellar UV/visible spectral diversity, thereby fully characterizing the multi-dimensional parameter space governing sub-Neptune atmospheric diversity. The diagnostic framework developed here provides a foundation for more sophisticated atmospheric interpretation methods essential for advancing our understanding of the most abundant class of planets in the Galaxy.

\begin{acknowledgments}

This work was partially supported by funding from the Center for Exoplanets and Habitable Worlds. The Center for Exoplanets and Habitable Worlds is supported by the Pennsylvania State University and the Eberly College of Science. Neha Dushyantha Kumar acknowledges funding support from the Erickson Discovery Grant for Summer 2025, which enabled the comprehensive parameter space exploration presented in this work. 

We thank Shang-Min Tsai for his invaluable assistance and guidance in implementing  \texttt{VULCAN} for our sub-Neptune atmospheric modeling, and for his contributions to the photochemical network development that made this systematic parameter space exploration possible. We also thank Luis Welbanks for his insightful discussions on K2-18b atmospheric characterization and for his presentation at the American Astronomical Society 247th meeting, which provided valuable perspective that helped shape the goals and direction of this manuscript.

The authors of this work recognize the (i) Penn State Institute for Computational and Data Sciences (RRID:SCR\_025154) for providing access to computational research infrastructure within the Roar Core Facility (RRID: SCR\_026424) and (ii) NASA High-End Computing (HEC) Program through the NASA Center for Climate Simulation (NCCS) at Goddard Space Flight Center.

CIC acknowledges support by NASA Headquarters through an appointment to the NASA Postdoctoral Program at the Goddard Space Flight Center, administered by ORAU through a contract with NASA, and support from NASA under award number 80GSFC24M0006.

\end{acknowledgments}

%


\software{  
\texttt{PICASO} \citep{Batalha2022, Mukherjee2023},
\texttt{VULCAN} \citep{Tsai2021}
\texttt{exoatlas} \citep{Berta-Thompson2025},
\texttt{matplotlib} \citep{4160265},
\texttt{numpy} \citep{5725236},
\texttt{pandas} \citep{article},
          }

\bibliographystyle{aasjournalv7}
\bibliography{main}

\appendix
\renewcommand{\thetable}{A\arabic{table}}
\setcounter{table}{0}

\renewcommand{\thefigure}{A\arabic{figure}}
\setcounter{figure}{0}

Table~\ref{tab:Abundance_trends} summarizes the molecular abundance ranges from our systematic parameter-space exploration, providing a reference for the chemical diversity across different thermal and dynamical regimes in K2-18b analog atmospheres. These abundance ranges, derived from our coupled \texttt{PICASO}--\texttt{VULCAN} modeling grid (Figures~\ref{fig:abundances-six} and \ref{fig:abundance_profiles_all}), can inform atmospheric retrievals, guide observational strategies, and facilitate model comparisons. The temperature-regime classifications reflect distinct chemical behaviors identified in Section~\ref{sec:Results}, while mixing dependencies highlight species-specific sensitivities that serve as atmospheric dynamics diagnostics. Values represent photospheric means accounting for coupled thermochemical, transport, and photochemical processes across our full parameter space.

\begin{table}[!h]
\centering
\begin{tabular}{|c|c|c|c|c|}
\hline
Molecule & Temperature Regime & Mixing Conditions & Abundance Range & $\log_{10}$ Abundance \\
\hline
\multirow{3}{*}{\textbf{CH$_4$ (Methane)}} 
  & Cool ($<150\,\mathrm{K}$) 
  & All $K_{\mathrm{zz}}$ ($10^{5}$--$10^{12}\,\mathrm{cm}^2\,\mathrm{s}^{-1}$) 
  & $10^{-3}$ to $10^{-2}$ & $-3$ to $-2$ \\
\cline{2-5}
  & Intermediate ($200$--$350\,\mathrm{K}$) 
  & All $K_{\mathrm{zz}}$ ($10^{5}$--$10^{12}\,\mathrm{cm}^2\,\mathrm{s}^{-1}$) 
  & $10^{-4}$ to $10^{-3}$ & $-4$ to $-3$ \\
\cline{2-5}
  & Hot ($>400\,\mathrm{K}$) 
  & All $K_{\mathrm{zz}}$ ($10^{5}$--$10^{12}\,\mathrm{cm}^2\,\mathrm{s}^{-1}$) 
  & $10^{-6}$ to $10^{-5}$ & $-6$ to $-5$ \\
\hline

\multirow{3}{*}{\textbf{NH$_3$ (Ammonia)}} 
  & Cool ($<150\,\mathrm{K}$) 
  & All $K_{\mathrm{zz}}$ ($10^{5}$--$10^{12}\,\mathrm{cm}^2\,\mathrm{s}^{-1}$) 
  & $10^{-3}$ to $10^{-2.5}$ & $-3$ to $-2.5$ \\
\cline{2-5}
  & Intermediate ($200$--$350\,\mathrm{K}$) 
  & All $K_{\mathrm{zz}}$ ($10^{5}$--$10^{12}\,\mathrm{cm}^2\,\mathrm{s}^{-1}$) 
  & $10^{-4}$ to $10^{-3}$ & $-4$ to $-3$ \\
\cline{2-5}
  & Hot ($>400\,\mathrm{K}$) 
  & All $K_{\mathrm{zz}}$ ($10^{5}$--$10^{12}\,\mathrm{cm}^2\,\mathrm{s}^{-1}$) 
  & $10^{-5.5}$ & $-5.5$ \\
\hline

\multirow{4}{*}{\textbf{CO (Carbon Monoxide)}} 
  & Cool ($<150\,\mathrm{K}$) 
  & Low $K_{\mathrm{zz}}$ ($10^{5}$--$10^{8}\,\mathrm{cm}^2\,\mathrm{s}^{-1}$) 
  & $10^{-3}$ to $10^{-2}$ & $-3$ to $-2$ \\
\cline{2-5}
  & Cool ($<150\,\mathrm{K}$) 
  & High $K_{\mathrm{zz}}$ ($10^{10}$--$10^{12}\,\mathrm{cm}^2\,\mathrm{s}^{-1}$) 
  & $10^{-2}$ to $10^{-1.5}$ & $-2$ to $-1.5$ \\
\cline{2-5}
  & Intermediate ($200$--$350\,\mathrm{K}$) 
  & All $K_{\mathrm{zz}}$ ($10^{5}$--$10^{12}\,\mathrm{cm}^2\,\mathrm{s}^{-1}$) 
  & $10^{-1.5}$ to $10^{-1}$ & $-1.5$ to $-1$ \\
\cline{2-5}
  & Hot ($>400\,\mathrm{K}$) 
  & All $K_{\mathrm{zz}}$ ($10^{5}$--$10^{12}\,\mathrm{cm}^2\,\mathrm{s}^{-1}$) 
  & $10^{-1}$ & $-1$ \\
\hline

\multirow{4}{*}{\textbf{CO$_2$ (Carbon Dioxide)}} 
  & Cool ($<150\,\mathrm{K}$) 
  & All $K_{\mathrm{zz}}$ ($10^{5}$--$10^{12}\,\mathrm{cm}^2\,\mathrm{s}^{-1}$) 
  & $10^{-4.5}$ to $10^{-3.5}$ & $-4.5$ to $-3.5$ \\
\cline{2-5}
  & Intermediate ($200$--$350\,\mathrm{K}$) 
  & Low $K_{\mathrm{zz}}$ ($10^{5}$--$10^{8}\,\mathrm{cm}^2\,\mathrm{s}^{-1}$) 
  & $10^{-2.5}$ to $10^{-2}$ & $-2.5$ to $-2$ \\
\cline{2-5}
  & Intermediate ($200$--$350\,\mathrm{K}$) 
  & High $K_{\mathrm{zz}}$ ($10^{10}$--$10^{12}\,\mathrm{cm}^2\,\mathrm{s}^{-1}$) 
  & $10^{-2}$ to $10^{-1.5}$ & $-2$ to $-1.5$ \\
\cline{2-5}
  & Hot ($>400\,\mathrm{K}$) 
  & All $K_{\mathrm{zz}}$ ($10^{5}$--$10^{12}\,\mathrm{cm}^2\,\mathrm{s}^{-1}$) 
  & $10^{-4.5}$ to $10^{-3.5}$ & $-4.5$ to $-3.5$ \\
\hline

\multirow{3}{*}{\textbf{H$_2$O (Water)}} 
  & Cool ($<150\,\mathrm{K}$) 
  & All $K_{\mathrm{zz}}$ ($10^{5}$--$10^{12}\,\mathrm{cm}^2\,\mathrm{s}^{-1}$) 
  & $10^{-3}$ to $10^{-2.5}$ & $-3$ to $-2.5$ \\
\cline{2-5}
  & Intermediate ($200$--$350\,\mathrm{K}$) 
  & All $K_{\mathrm{zz}}$ ($10^{5}$--$10^{12}\,\mathrm{cm}^2\,\mathrm{s}^{-1}$) 
  & $10^{-1.5}$ to $10^{-1}$ & $-1.5$ to $-1$ \\
\cline{2-5}
  & Hot ($>400\,\mathrm{K}$) 
  & All $K_{\mathrm{zz}}$ ($10^{5}$--$10^{12}\,\mathrm{cm}^2\,\mathrm{s}^{-1}$) 
  & $10^{-1}$ to $10^{-0.5}$ & $-1$ to $-0.5$ \\
\hline

\multirow{5}{*}{\textbf{HCN (Hydrogen Cyanide)}} 
  & Cool ($<150\,\mathrm{K}$) 
  & All $K_{\mathrm{zz}}$ ($10^{5}$--$10^{12}\,\mathrm{cm}^2\,\mathrm{s}^{-1}$) 
  & $10^{-8.5}$ to $10^{-8}$ & $-8.5$ to $-8$ \\
\cline{2-5}
  & Intermediate ($200$--$350\,\mathrm{K}$) 
  & Low $K_{\mathrm{zz}}$ ($10^{5}$--$10^{8}\,\mathrm{cm}^2\,\mathrm{s}^{-1}$) 
  & $10^{-7}$ to $10^{-6}$ & $-7$ to $-6$ \\
\cline{2-5}
  & Intermediate ($200$--$350\,\mathrm{K}$) 
  & High $K_{\mathrm{zz}}$ ($>10^{10}\,\mathrm{cm}^2\,\mathrm{s}^{-1}$) 
  & $10^{-8.5}$ to $10^{-8}$ & $-8.5$ to $-8$ \\
\cline{2-5}
  & Hot ($>400\,\mathrm{K}$) 
  & Low $K_{\mathrm{zz}}$ ($10^{5}\,\mathrm{cm}^2\,\mathrm{s}^{-1}$) 
  & $10^{-5.5}$ to $10^{-5}$ & $-5.5$ to $-5$ \\
\cline{2-5}
  & Hot ($>400\,\mathrm{K}$) 
  & High $K_{\mathrm{zz}}$ ($>10^{10}\,\mathrm{cm}^2\,\mathrm{s}^{-1}$) 
  & $10^{-8.5}$ to $10^{-8}$ & $-8.5$ to $-8$ \\
\hline
\end{tabular}
\caption{Molecular abundance ranges in K2-18b analog atmospheres. Mean molecular abundances (volume mixing ratios) in the photospheric region ($10^{-3}$--$10^{-4}$~bar) as a function of intrinsic temperature and vertical mixing strength. Results from coupled PICASO--VULCAN atmospheric models spanning $T_{\mathrm{int}} = 60$--$500$~K and $K_{zz} = 10^{5}$--$10^{12}\,\mathrm{cm}^2\,\mathrm{s}^{-1}$. Temperature regimes: cool ($<150$~K), intermediate (200--350~K), and high ($>400$~K).}
\label{tab:Abundance_trends}
\end{table}

\begin{figure}[h]
    \centering

    \begin{subfigure}{0.95\linewidth}
        \centering
        \includegraphics[width=\linewidth]{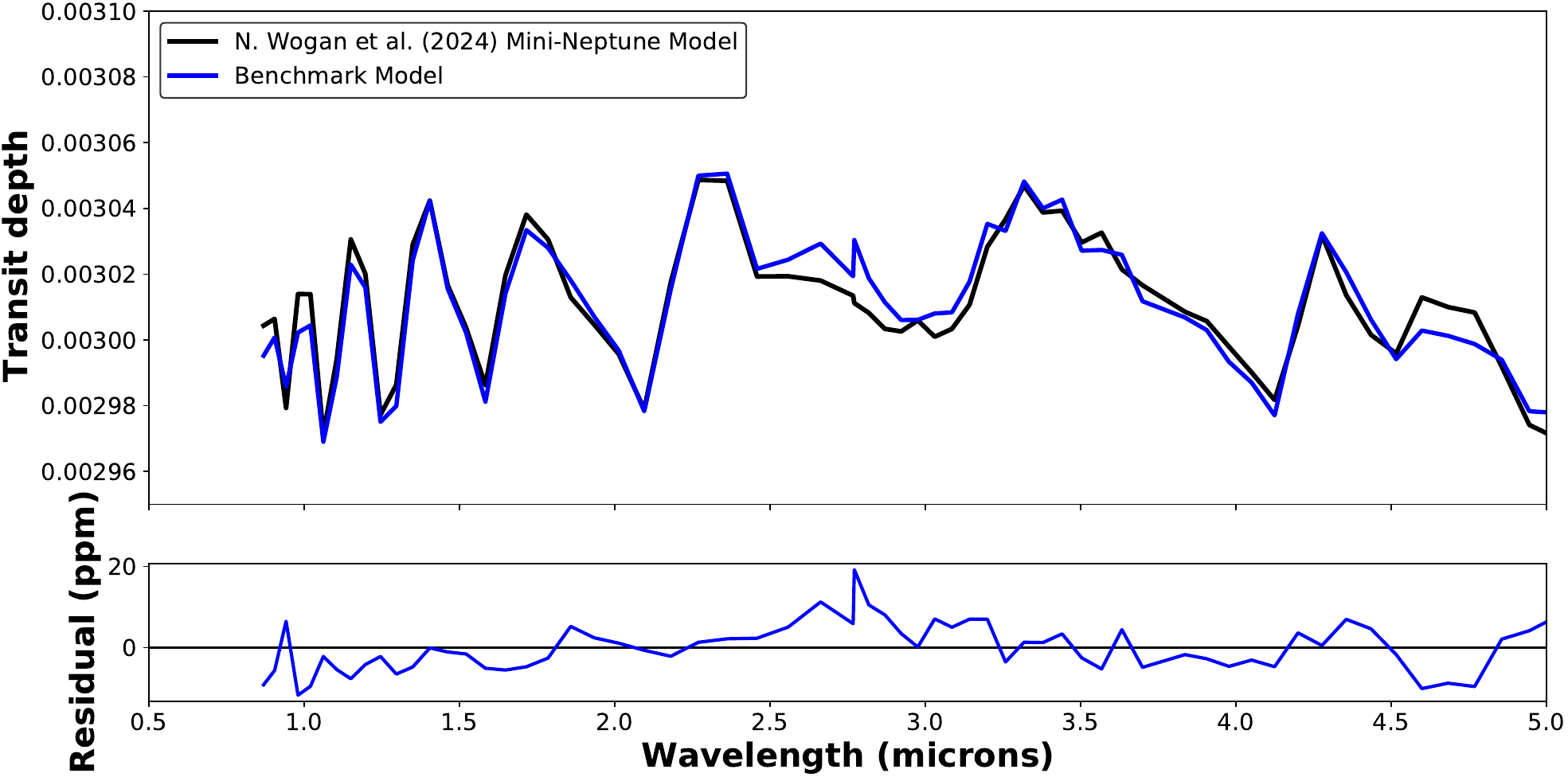}
        \captionsetup{skip=10pt}
        \caption{N. Wogan et al. (2024) Mini-Neptune Model vs. Benchmark Model}
        \label{fig:plot1}
    \end{subfigure}

    \vspace{0.6em} 

    \begin{subfigure}{0.95\linewidth}
        \centering
        \includegraphics[width=\linewidth]{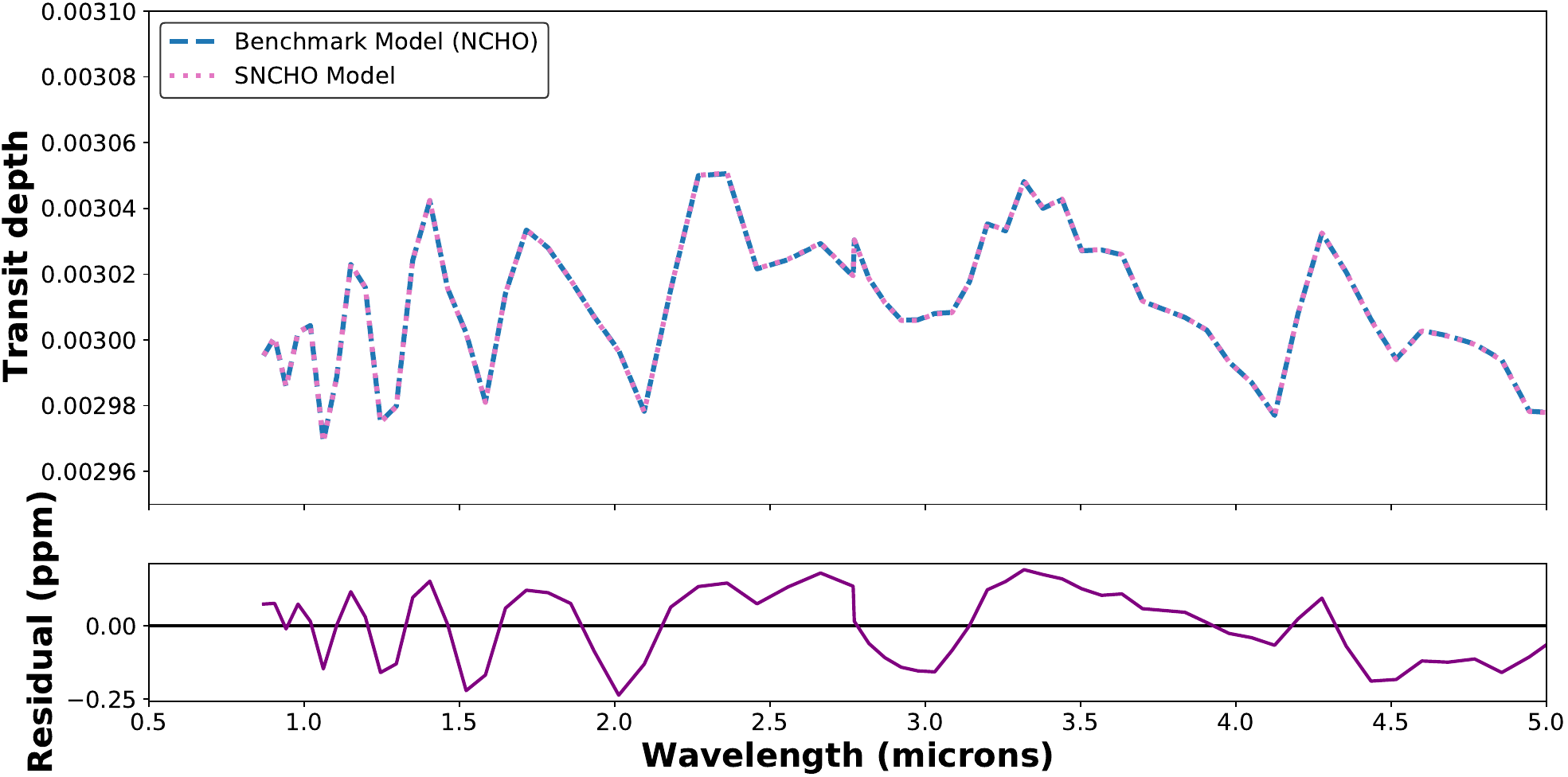}
        \captionsetup{skip=10pt}
        \caption{Benchmark vs. SNCHO Model}
        \label{fig:plot2}
    \end{subfigure}

    \caption{Comparison plots for K2-18b model spectra. Panel~(a) shows the transmission spectrum comparison between the N.\ Wogan et~al.\ (2024) Mini-Neptune Model (black line) and our Benchmark Model (blue line) across wavelengths from $0.5$--$5.0~\mu\mathrm{m}$, with residuals plotted below showing differences typically $<10~\mathrm{ppm}$ except in the $2.5$--$3.0~\mu\mathrm{m}$ region. Panel~(b) shows the comparison between the Benchmark model (blue dashed line) and our SNCHO model (pink doted line), demonstrating the minimal impact of sulfur chemistry inclusion on the primary spectral features. The excellent agreement validates our \texttt{PICASO-VULCAN} modeling framework, with wavelength-averaged differences of $1.2 \pm 0.5~\mathrm{ppm}$ between NCHO and SNCHO models, confirming that sulfur chemistry exclusion does not introduce systematic biases in our key diagnostic species (CH$_4$, CO$_2$, H$_2$O, NH$_3$).}
    \label{fig:stacked_two}
\end{figure}

Figure \ref{fig:stacked_two} demonstrates the robustness of our atmospheric modeling approach through two key validation tests. First, we benchmark our \texttt{PICASO-VULCAN} framework against the established \citet{Wogan2024} mini-Neptune model for K2-18b, showing excellent agreement across most wavelengths with differences typically below observational precision. The largest systematic offset occurs in the $2.5$--$3.0~\mu\mathrm{m}$ region ($\sim 15.7~\mathrm{ppm}$). Second, we validate our decision to use the simplified NCHO chemical network by comparing against the full SNCHO network, finding negligible differences ($1.2 \pm 0.5~\mathrm{ppm}$ wavelength-averaged) that confirm sulfur chemistry does not significantly impact our primary diagnostic molecules. While this validation supports our current approach, future work should systematically explore the full parameter grid using the SNCHO chemical network to comprehensively assess potential sulfur chemistry effects across the complete range of intrinsic temperatures and vertical mixing coefficients. These validation tests support the reliability of our systematic parameter space exploration and strengthen confidence in our conclusions about the role of intrinsic temperature and vertical mixing in sub-Neptune atmospheric composition.

Table~\ref{tab:quench_timescales_side_by_side} quantifies the critical timescales and pressure levels governing disequilibrium chemistry in K2-18b analog atmospheres, providing diagnostics for how interior thermal states and atmospheric mixing determine observable molecular abundances \citep{Visscher2006, Visscher2010, Moses2011}. These quenching parameters define transition points where rapid vertical transport freezes deep atmospheric compositions that are transported to the observable photosphere \citep{Visscher2010}. The systematic variation of chemical timescales with $K_{\mathrm{zz}}$ and temperature reveals the physical mechanisms underlying our abundance trends (Table~\ref{tab:Abundance_trends}) and explains why different molecules exhibit distinct thermal and dynamical sensitivities \citep{Moses2011, Hu2021}. Comparison between variable and constant $K_{\mathrm{zz}}$ profiles demonstrates that altitude-dependent mixing creates more complex quenching behavior than simple parameterizations \citep{Wogan2024}, highlighting the importance of realistic transport modeling for accurate atmospheric predictions

\begin{table}[h]
\centering

\begin{subtable}[t]{0.49\textwidth}
\centering
\caption{Methane (CH$_4$)}
\label{tab:CH4_quench_main}
\begin{tabular}{|c|c|c|c|}
\toprule
$K_{\mathrm{zz}}$ (cm$^{2}$ s$^{-1}$) & $T_q$ (K) & $P_q$ range (bar) & $\tau_{\mathrm{chem}}$ (s) \\
\midrule
\multicolumn{4}{|c|}{\textbf{Variable $K_{\mathrm{zz}}$}} \\
\midrule
$1\times10^{5}$   & 60  & $10^{1}$--$10^{2}$  & $123 \pm 921$ \\
$1\times10^{5}$   & 250 & $10^{0}$--$10^{1}$  & $2142 \pm 15996$ \\
$5.6\times10^{6}$ & 450 & $10^{-1}$--$10^{0}$ & $149 \pm 1115$ \\
\midrule
\multicolumn{4}{|c|}{\textbf{Constant $K_{\mathrm{zz}}$}} \\
\midrule
$1\times10^{5}$   & 60  & $10^{1}$--$10^{2}$  & $1.5 \pm 11.1$ \\
$1\times10^{5}$   & 250 & $10^{0}$--$10^{1}$  & $25.9 \pm 193.5$ \\
$1\times10^{5}$   & 450 & $10^{-1}$--$10^{0}$ & $84.0 \pm 627.1$ \\
\midrule
$1\times10^{8}$   & 60  & $10^{1}$--$10^{2}$  & $0.242 \pm 1.806$ \\
$1\times10^{8}$   & 250 & $10^{0}$--$10^{1}$  & $4.2 \pm 31.4$ \\
$1\times10^{8}$   & 450 & $10^{-1}$--$10^{0}$ & $13.6 \pm 101.6$ \\
\midrule
$1\times10^{10}$  & 60  & $10^{2}$--$10^{2.5}$ & $0.024 \pm 0.181$ \\
$1\times10^{10}$  & 250 & $10^{1}$--$10^{1.5}$ & $0.420 \pm 3.135$ \\
$1\times10^{10}$  & 450 & $10^{0}$--$10^{1}$   & $1.4 \pm 10.2$ \\
\midrule
$1\times10^{12}$  & 60  & $10^{2}$--$10^{2.5}$ & $(4.0 \pm 29.9)\times10^{-3}$ \\
$1\times10^{12}$  & 250 & $10^{1}$--$10^{2}$   & $0.069 \pm 0.518$ \\
$1\times10^{12}$  & 450 & $10^{0}$--$10^{1}$   & $0.225 \pm 1.679$ \\
\bottomrule
\end{tabular}
\end{subtable}
\hspace{0.009\textwidth} 
\begin{subtable}[t]{0.49\textwidth}
\centering
\caption{Ammonia (NH$_3$)}
\label{tab:NH3_quench_main}
\begin{tabular}{|c|c|c|c|}
\toprule
$K_{\mathrm{zz}}$ (cm$^{2}$ s$^{-1}$) & $T_q$ (K) & $P_q$ range (bar) & $\tau_{\mathrm{chem}}$ (s) \\
\midrule
\multicolumn{4}{|c|}{\textbf{Variable $K_{\mathrm{zz}}$}} \\
\midrule
$1\times10^{5}$  & 60  & $10^{2}$            & $100.0 \pm 746.3$ \\
$1\times10^{5}$  & 250 & $10^{1}$            & $1735 \pm 12957$ \\
$1\times10^{5}$  & 450 & $10^{0}$--$10^{1}$  & $5622 \pm 41979$ \\
\midrule
\multicolumn{4}{|c|}{\textbf{Constant $K_{\mathrm{zz}}$}} \\
\midrule
$1\times10^{5}$  & 60  & $10^{2}$            & $1.2 \pm 9.0$ \\
$1\times10^{5}$  & 250 & $10^{1}$            & $21.0 \pm 156.8$ \\
$1\times10^{5}$  & 450 & $10^{0}$--$10^{1}$  & $68.0 \pm 507.9$ \\
\midrule
$1\times10^{8}$  & 60  & $10^{2}$            & $0.196 \pm 1.463$ \\
$1\times10^{8}$  & 250 & $10^{1}$            & $3.4 \pm 25.4$ \\
$1\times10^{8}$  & 450 & $10^{0}$--$10^{1}$  & $11.0 \pm 82.3$ \\
\midrule
$1\times10^{10}$ & 60  & $10^{2}$            & $0.020 \pm 0.146$ \\
$1\times10^{10}$ & 250 & $10^{1}$            & $0.340 \pm 2.539$ \\
$1\times10^{10}$ & 450 & $10^{1}$            & $1.1 \pm 8.2$ \\
\midrule
$1\times10^{12}$ & 60  & $10^{2}$            & $(3.2 \pm 24.2)\times10^{-3}$ \\
$1\times10^{12}$ & 250 & $10^{1}$            & $0.056 \pm 0.420$ \\
$1\times10^{12}$ & 450 & $10^{1}$            & $0.182 \pm 1.360$ \\
\bottomrule
\end{tabular}
\end{subtable}
\caption{Quenching timescales for CH$_4$ and NH$_3$ as a function of vertical eddy diffusion coefficient ($K_{\mathrm{zz}}$). Quench pressure indicates the atmospheric level where chemical and mixing timescales become equal ($\tau_{\mathrm{chem}}=\tau_{\mathrm{mix}}$).}
\label{tab:quench_timescales_side_by_side}

\end{table}

\end{document}